\documentclass[usenatbib,twocolumn]{mn2e}
\usepackage{graphicx,amsmath,amssymb,enumitem,verbatim}
\begin{document}

\title{Some observational tests of a minimal galaxy formation model}

\author[Observational tests of a minimal model]{J. D. Cohn${}^{1,2}$\thanks{E-mail: jcohn@berkeley.edu}\\
${}^1$Space Sciences Laboratory 
  University of California, Berkeley, CA 94720, USA\\
${}^2$Theoretical Astrophysics Center,
  University of California, Berkeley, CA 94720, USA}
\maketitle
\begin{abstract}
Dark matter simulations can serve as a basis for creating galaxy
histories via the galaxy-dark matter connection.  
Here, one such model by \citet{Bec15} is implemented with several
variations on three
different
dark matter simulations.  Stellar mass and
star formation rates are assigned to all simulation subhalos at all times,
using subhalo mass gain to determine stellar mass gain.
The observational properties of the resulting galaxy distributions 
are compared to each other and
observations for a range of redshifts from 0-2. 
Although many of the galaxy distributions seem reasonable, there are
noticeable differences as simulations, subhalo mass gain definitions, or subhalo mass definitions are altered,
suggesting that the model should change as these properties are
varied.   Agreement with observations may improve by including
redshift dependence in the
added-by-hand random contribution to star formation rate.  
There appears to be an excess of faint quiescent
galaxies as well (perhaps due in part to
differing definitions of quiescence).
The ensemble of galaxy formation
histories for these models
tend to have more scatter around
their average histories (for a fixed final stellar mass) than the two more 
predictive and
elaborate semi-analytic models of \citet{Guo13,Hen15}, and require
more basis fluctuations (using PCA) to capture 90\% of the scatter
around their average histories.

The codes to plot model predictions (in some cases alongside
observational data) are publicly 
available to test other mock catalogues at
https://github.com/jdcphysics/validation/ .
Information on how to use these codes is in the appendix.  
\end{abstract}
\section{Introduction}
Galaxies are expected to form within the deep potential wells of dark
matter halos (\citet{WhiRee78,Blu84}, for a general introduction see
\citet{MovdBWhi10}).  This galaxy-dark matter connection suggests that
simulations of the histories and spatial distributions of dark matter
halos can be used as scaffolding for models of galaxy histories and
distributions.  Simulating galaxy properties based upon dark matter simulations 
ranges from from detailed semi-analytic models (see, for example, reviews by 
\citet{Bau06,Ben10}) of several galaxy
properties at all times, which model and predict many different
processes, to models such as the halo model which assign galaxies of a
certain kind to dark matter halos at a fixed time by requiring that
they match observed clustering and number counts
(\citet{Sel00,PeaSmi00,CooShe02}, also see more recent incarnations
such as \citet{Hea16}) or variants such as conditional luminosity
functions \citep{YanMovdB03}) or abundance matching in luminosity and
(proxy for) halo mass
(e.g. \citet{ValOst06,ConWec09}).  The resulting 
galaxy distributions can then be compared to observations, testing the
physical assumptions used in their construction.  These methods are also used to construct mock galaxy catalogues (synthetic skies) to
aid in designing and analyzing galaxy surveys.  More time intensive
hydrodynamical simulations, which include and thus fix the baryonic
physics and subgrid models for each cosmological run are also being
developed, see e.g. \citet{BorKra12,Nei12} for some comparisons of
trade-offs.  All of these approaches are currently being developed
 and extended. 

Here, a simple model defined by \citet{Bec15}\footnote{\citet{Bec15} more
  generally gives a probabilistic framework for 
combining simulations and observations to
get simulated galaxy properties consistent with the chosen observations.} is
explored.
It 
creates statistical
samples of galaxy formation
histories based upon the growth of dark matter subhalos,
producing stellar masses and star formation rates for every subhalo
in the simulation.  (In what follows the terms subhalo and halo will be
used interchangeably unless specifically noted.)
Much of the physics is encapsulated in
an average relation between stellar mass and halo growth found by 
\citet{BWCz8,BWCshort}.  \citet{BWCz8} matched observations of stellar
mass functions and star formation rates to average dark matter halo histories
at a series of redshifts, and found that 
\begin{equation}
\frac{d M^*}{dt} \sim f(M_h,z(t)) \frac{dM_h}{dt} \; .
\label{eq:dmsdmh}
\end{equation}  
Each subhalo has stellar mass $M^*$ and subhalo mass $M_h$ (virial
mass, in their approach).
The star formation efficiency
$f(M_h, z(t))$ is a weakly time dependent function
of subhalo mass $M_h$ and is publicly available at
www.peterbehroozi.com/data.html.

The simple \citet{Bec15} model uses this
relation for average stellar mass gain to
assign stellar mass and star formation rates to individual subhalos
throughout their histories
in a dark matter simulation, once a rule for inheriting stellar mass
from progenitor galaxies is added.  
\citet{Bec15} suggested one such rule for inheriting stellar mass,
as well as the addition of a random component to star formation.  
(Stellar mass is also lost due to aging.)  
Galaxy distributions produced by the \citet{Bec15} model, with his 
simulation,
have a $z=0$ galaxy stellar mass function close to observations.  The
$z=0$ star formation rates are also bimodal,
although not agreeing as closely with observations in detail \citep{Bec15}.

A wealth of galaxy properties follow from having a stellar mass and
star formation rate attached to each galaxy throughout its history,
in addition to the stellar masses and star formation rates themselves.
Colors
can be found by integrating a stellar population synthesis model over the
star formation rate history.  Galaxy positions, velocities and
environments are immediate, inherited from the host dark matter simulation.
In particular, colors are linked to environment inasmuch as
environment affects halo growth.
It is thus interesting to examine further properties of this model
beyond redshift zero,
and how it depends upon different simulations and other properties.
The inheritance of stellar mass, the definitions of halo mass and halo
mass gain, and the underlying dark matter simulation are all varied
here, and compared to observations at redshifts 0 to 2.
In addition, properties of the ensemble of resulting galaxy formation histories
 are compared to those of three other methods, two full blown
semi-analytic models and one straw man model.

Several other simple models have also been proposed, predicting
an assortment of galaxy properties for redshifts zero and above, for example
those by
\citet{Wan07,Bou10,Cat11,MutCroPoo13,Lil13,TacTreCar13,Bir14,Lu14,Lu15}.
The approach in the \citet{Bec15} model
seems closest to that of
\citet{MutCroPoo13}, as the stellar masses are tied to subhalo
properties directly (with the gas physics implicit). 

The underlying simulations and the
construction of the galaxy histories are described in \S
\ref{sec:background}.  In \S \ref{sec:simmeas}, comparisons are made
with several observations at different redshifts.
The ensemble of galaxy histories of this simple model are
compared via PCA (as in \citet{CohVdV15}) to some other more predictive models based upon
dark matter simulations, in \S \ref{sec:histories}.  Discussion and
conclusions are in \S\ref{sec:conc}.
The appendix \S\ref{sec:appmsmh} 
gives the formulae in detail for the two stellar mass
to halo mass prescriptions which are compared to the models in \S \ref{sec:simmeas}.

Appendix \S\ref{sec:howto} describes  
how to make the plots used in \S\ref{sec:simmeas}
for a single mock galaxy catalogue, using codes at
https://www.github.com/jdcphysics/validation/ (the
code valid\_suite.py in the code subdirectory vsuite).  To use this
code, a list of galaxy
stellar masses, star formation rates and subhalo masses are needed
as input, which
can lie in either a light cone, a fixed time periodic box, or just
have some fixed redshift and a random position (if generated analytically).
Most of the observational data for these tests are in the redshift
range 0 to 1, i.e. overlapping with redshifts at the centers of the currently
running dark energy survey (DES, www.darkenergysurvey.org)
and the upcoming LSST (www.lsst.org).

\section{Method for making galaxy histories}
\label{sec:background}
\subsection{Dark matter histories}
The galaxy histories in \citet{Bec15} are
 based upon subhalo histories
in a particular dark matter simulation.  Three different simulations are
considered here, with parameters and other details shown in Table
\ref{tab:dmsims}, along with the names and properties of the different
variations constructed with the simulations.  Adding a ``2'' in front
of the names
listed in Table \ref{tab:dmsims} refers to the
inheritance of the stellar mass from the two most massive progenitors
rather than just the most massive progenitor, described in \S\ref{sec:construction} below.
The simulation used by \citet{Bec15} is also listed for comparison.
All of the simulations were done with 
periodic boxes, with sides of length $\sim 250 h^{-1}
Mpc$.  They differ in
time resolution, particle number, cosmological parameters, methods for
making trees and methods for calculating halo masses. 

\begin{table*}
\centering 
\begin{tabular}{l|l|c|c|c|c|c|c|c|c|c|c|} 
Name&Sim&particles & $\Omega_m$ 
& $\Omega_b h^2$&$\sigma_8$ & n&h&
$M_{\rm halo}$&$\Delta M_h$&trees\\ \hline
tree&TreePM &2560${}^3$&0.3085&0.022&0.83&0.9611&0.6777&$M_{\rm
                                                         fof}$&all M&
\citet{StaWhiLee15}\\
  \hline
bolp&Bolshoi-P&2048${}^3$&0.307&0.022&0.823&0.96&0.678&$SAM\_M_{\rm
                                                             vir}$&Mmp&Rockstar\\
bolpmvir& & & & & & & & $M_{\rm vir}$ &all M& \\
bolpmmp& & & & & & & & $M_{\rm vir}$ &Mmp& \\
  \hline
bol&Bolshoi&2048${}^3$&0.270&0.023&0.820&0.95&0.70&$SAM\_M_{\rm
                                                                vir}$&Mmp                                                        
                                                        &Rockstar\\
bolmvir&  & & & & & && $M_{\rm  vir}$ &all M&\\
bolmmp&  & & & & & && $M_{\rm  vir}$ &Mmp&\\
\hline 
orig&\citet{Bec15} &2048${}^3$&0.286&0.0235&0.82&0.96&0.70&$M_{\rm
                                                              vir}$ &Mmp&Rockstar \\
  \hline
\end{tabular} %
\caption{Properties of the three dark matter simulations used here, and the
  one originally used to implement the \citet{Bec15} model.
 They
  have similar volume (250 $h^{-1}$Mpc sides for all except TreePM, which
  has a 256 $h^{-1}$ Mpc side) but differ in time and mass
  resolution,  
cosmology, tree construction and halo mass definition.
The names for the variants will be used below.
The bolp and bol variants use $SAM\_M_{\rm vir}$, the instantaneous $M_{\rm vir}$ boxcar smoothed over the
current, past and future steps (the sum divided by 3) and then combined with
the requirement that a descendant
mass is never less than the sum of its progenitors (this mass is
provided by Rockstar, described in \citet{ctrees,rockstar}). 
Accreted mass
is found using Eq.~\ref{eq:samacmass}, ``Mmp'', for $SAM\_M_{\rm vir}$
(bol and
bolp variants) and instantaneous Rockstar mass
$M_{\rm vir}$ 
(bolpmmp, bolmmp and the original \citet{Bec15} model.
Accreted mass is calculated using Eq.~\ref{eq:nosam} for
 bolmvir and bolpmvir (with instantaneous 
virial masses $M_{\rm vir}$), and tree (with instantaneous
$M_{\rm fof}$).  Adding a ``2'' in front of the name refers to the
inheritance of the stellar mass from the two most massive progenitors
rather than just the most massive progenitor, described in \S\ref{sec:construction} below.
The TreePM simulation is described in \citet{StaWhiLee15}, Bolshoi-P
is described in
\citet{Kly16,BolPla16}, and Bolshoi is described in \citet{KlyTruPri11}.  The tree
constructions are described in \citet{StaWhiLee15} for TreePM and
\citet{ctrees,rockstar} (named Rockstar) for the other runs.
}
\label{tab:dmsims} 
\end{table*}

In more detail,
the TreePM dark matter simulation was created with the TREEPM
\citep{TreePM} code, in a periodic box of side $ 256 h^{-1}
Mpc$.  
This simulation is
described in more detail in \citet{StaWhiLee15}.  It has 128 outputs 
between redshift 0 and 10, evenly spaced in $\ln a $, the scale
factor.
Halos are found using the Friends of friends (FoF) algorithm
\citep{DEFW} with linking length $b=$ 0.168.  
Galaxies inhabit simulation subhalos.
A central subhalo is the whole FoF halo and a satellite subhalo is the
remnant of an FoF halo which has fallen into a larger FoF halo.  Satellites 
are found and tracked using a 6-dimensional phase
space finder \citep{DieKuhMad06}, FoF6d, along with tracking between
times and skipped outputs
as described in \citet{WhiCohSmi10,WetWhi10}.  Satellite subhalo masses are taken to
be their infall FoF halo mass,
which can increase only if a merger occurs, and a satellite is removed when its FoF6d mass
falls below 0.007 of its infall mass as in \citet{WetWhi10}. 
Phantom halos (with no progenitor) sometimes appear, presumably
  these are splashback halos, satellites which travelled out of a larger
  halo far enough to be recognized as a separate halo, rather than
  remaining a satellite within the larger halo.  Phantoms 
which appear at the last time step are discarded.

  The other two simulations used here are the original Bolshoi simulation
  \citep{KlyTruPri11}, and its updated counterpart with \citet{Pla13}
  cosmological parameters, Bolshoi-P \citep{Kly16,BolPla16}.
Both simulations are run with ART. These have higher
  particle mass and slightly smaller box size ($250 h^{-1}$Mpc) 
than the TreePM
  simulation.  The two Bolshoi, Bolshoi-P simulations
have finer time resolution at late time (after $z\sim 2$) relative to
TREEPM.  Bolshoi (Bolshoi-P) has in total 181 (178)
  steps from redshift $\sim$ 14  (17), spaced in
  steps of constant $a$.  The spacing for Bolshoi changes from
$\Delta a = 0.003$ to twice that at
 $a<0.6834$, for Bolshoi-P the spacing changes from 0.0076 for
 $a>0.8084$ to 0.005 at earlier times.
A few output $a$ values seem to be skipped in the Bolshoi simulation.
The Bolshoi cosmology
  differs from that of TreePM, Bolshoi-P, and slightly from the
  original \citet{Bec15} simulation, as can be seen in Table
  \ref{tab:dmsims}.
The subhalo trees and mass histories are found using the
  methods of \citet{ctrees,rockstar}, more detail can be found in
  those papers.
  The dark matter halo $SAM\_M_{\rm vir}$ masses produced by
  \citet{ctrees,rockstar} are the virial halo masses $M_{\rm vir}$,
  boxcar smoothed over 3 time steps (along most massive progenitors), 
with the present step in the
  middle, and with
each central subhalo constrained
  to have mass at least as large as the sum of its
  progenitors. \footnote{See code at
    https://bitbucket.org/pbehroozi/consistent-trees/overview for
    details.} 

The original \citet{Bec15} simulation, to which the model 
parameters were tuned, is
closest to Bolshoi in cosmological parameters.  It has 100 time steps
equally separated in $\ln a$ between $z=0$ and $z=12$, i.e. with
time step separation slightly larger than TreePM.

  The simulations differ in
  cosmology, subhalo mass (the non-decreasing and
  smoothed $SAM\_M_{\rm vir}$ 
  vs. the instantaneous $M_{\rm fof}$ or $M_{\rm vir}$ which can increase
  or decrease at each time step), 
the definition of satellite masses \footnote{Subhalo mass definitions for
    the Rockstar trees are explained in detail in \citet{rockstar}:
    the masses are from an $M_{\rm vir}$ spherical overdensity
    calculation on the particles identified as subhalo members
    (using the phase space based friends of friend finder for both the
    subhalo and its host halo).   Properties of the mass
    gain and loss for
    satellites using this finder are studied in \citet{vdb16}.  The analogous
    $M_{\rm fof}$ for a satellite subhalo is 
  difficult to define in TreePM; their subhalo
  masses are set to their infall FoF masses and only increase due to mergers.}, tree construction, and the
  difference in time step spacing (shorter steps at later times in the
  Bolshoi, Bolshoi-P simulation relative to the TreePM and \citet{Bec15} simulations).
The chosen method to calculate accreted mass is
also given, both are from
\citet{Bec15}, and are defined in Eqs.~\ref{eq:samacmass} (Mmp) and Eq.~\ref{eq:nosam}
(all M) below.

\subsection{Creating galaxy stellar mass and star formation rate histories}
\label{sec:construction}
To get stellar masses and star formation rates at each output time,
the simulated dark matter subhalo histories described above are
tracked through time.  They are assigned
stellar mass and star formation rates as described in
\citet{Bec15}, except for the inheritance of stellar mass (see below):
\begin{itemize}
\item At each output time, the accreted halo mass of a galaxy is taken to be
  the
difference between its halo mass at that step 
and the weighted sum of the halo masses
of its progenitors at the previous time step.

The weights are motivated by the mass definitions of the simulation.  
For $SAM\_M_{\rm vir}$ masses, and for
the Rockstar $M_{\rm vir}$ masses in some cases  (including variants here
and in \citet{Bec15},\footnote{For \citet{Bec15}, this choice was motivated by the details of the
  construction of the star formation efficiencies, which used the most
  massive progenitor.}, 
only the subhalo mass of the most massive progenitor ($M_{\rm mmp}$) is
subtracted to get the accreted mass of the final subhalo,
\begin{equation}
\begin{array}{ll}
 \; \; \; &{\rm Mmp} \\
& \\
\Delta M_h(t) &= M_h(t) - M_{\rm mmp}(t_{\rm prev}) \; . \\
\end{array}
\label{eq:samacmass}
\end{equation}
For the TreePM box, with instantaneous $M_{\rm fof}$ masses, 
and for one pair of runs each for the Bolshoi(-P) runs, with the instantaneous $M_{\rm vir}$ masses, 
the formula suggested for including all progenitor masses 
in \citet{Bec15} is used
instead:
\begin{equation}
\begin{array}{lll}
\; \; \; && {\rm all \; M} \\
&& \\
\Delta M_h (t)&=& M_h (t) - M_{\rm mmp}(t_{\rm prev}) \\
& &- \frac{\sum M_{\rm
                not \;  mmp}(t_{\rm prev})
}{M_{\rm mmp}(t_{\rm prev})+\sum M_{\rm not \;  mmp}(t_{\rm prev})} M_h(t) \; . \\
\end{array}
\label{eq:nosam}
\end{equation}
If $\Delta M_h(t)$ is negative by the above, it is taken to be zero.
In both, $t$ is the present time and $t_{\rm prev}$ is the previous
time step output.
\item The change in stellar mass in this step has two contributions.
\begin{equation}
\Delta M^* = SFE(M_h,z) \Delta M_h f_b + \Delta M_{*,{\rm ran}}
\label{eq:dmstardmh}
\end{equation}
  One
contribution is from
the star formation efficiency found by \citet{BWCz8} as a
function of halo mass, times the change in halo mass times
the baryon fraction ($f_b=\Omega_b/\Omega_m$).  
That is, $f(M_h,z)$ in Eq. ~\ref{eq:dmsdmh} is taken to be $SFE(M_h,z) f_b$.
In practice, the time step between two simulation outputs 
is divided up into 250 time steps and
the halo mass is assumed to change the same amount between each.
The star formation efficiencies are tabulated at
http://www.peterbehroozi.com/data.html as the file sfe.dat in the
download tarball sfh\_z0\_z8.tar.gz .  The version release-sfr\_z0\_z8\_052913
is used in this paper.\footnote{For halo masses or times outside
of the range of these efficiencies, the stellar mass is placed on the
$M^*(M_h,z)$ relation from \citet{BWCz8}.
In contrast, \citet{Bec15} started galaxies at redshift 4 on the stellar mass-halo
mass relation.}  

The second contribution is
 a random
addition to the star formation rate given in \citet{Bec15}, times the
time step.
The random component to the star formation rate $ \dot{M}_{*,{\rm ran}}$ 
is drawn by taking a star
formation rate from a lognormal
distribution \footnote{As this relation was found in
  \citet{Bec15} for time steps which are different from those in the
 3 simulations used here, the scatter is likely only approximate for a
 general simulation.} of mean $\log_{10}(10^{-12}$yr${}^{-1}M^*)$ and scatter
$\epsilon =  0.25$.  The increase in stellar mass from this random
component is $\Delta M{_*,{\rm ran}} = \dot{M}_{*,{\rm ran}}(t-t_{\rm
  prev}) $. 

\item As implied by Eq.~\ref{eq:dmstardmh}, the 
total star formation rate is the
random star formation rate plus the star formation rate due to 
subhalo mass accretion at the
  last (small, 1/250th) time step.

 Without the random component, the galaxies with no subhalo mass gain
would have zero star formation rate.  The majority of the galaxies
with star formation only due to this random contribution to star formation rate
become quiescent galaxies below.\footnote{Thanks to M. Becker for explaining this
  point.}  
One way of viewing the whole construction is 
that galaxies whose subhalos gain
mass are roughly put on the star forming main sequence, with scatter
due to scatter in subhalo mass gain history, while those which do not
gain subhalo mass are quiescent, with some residual star formation
inserted by hand.

\item For a halo which has just appeared, the resulting stellar mass
  is the total stellar mass.

\item For a halo which has at least one progenitor in the simulation, the stellar
  mass at the current time is the newly formed stellar mass calculated
  above, plus 
  the stellar mass, after aging (below), of the most massive
  progenitor or of the two most massive progenitors.  Each run uses
  either one rule or the other for all galaxies and times; tree and
  2tree (see Table \ref{tab:dmsims} for properties of the tree model) for
  instance denote inheriting stellar mass from
 1 or 2 of the most massive progenitors, respectively.

  The rule for inheriting stellar mass is the main difference between
  the \citet{Bec15} model as implemented here and in the original
  paper.  
In \citet{Bec15}, the stellar mass of
  either the most massive or two most massive progenitors is given to
  a descendant galaxy depending upon whether there are 2 or more than
  2 progenitors.  This criterion is time step and resolution
  dependent, as the number of progenitors tends to grow as the time
  length between outputs grows, or as smaller halos are resolved.  As
  the time steps vary significantly between simulations considered here, both cases
  were instead considered separately.

\item Aging decreases the stellar mass inherited from the progenitors.  
The fraction of stellar mass lost by time $t$ since formation time $t_f$ is
taken to be \citep{BWCz8,MosNaaWhi13}
\begin{equation}
f_{\rm loss}(t-t_f) = 0.05  \ln( \frac{t-t_f}{1.4Myr} + 1)  \; .
\end{equation} 
For a given discrete time step, stellar mass loss due to aging is taken 
to be the average over the
interval that stars formed (between $t_{f1}$ and $t_{f2}$), 
as in \citet{BWCz8}, Eq.~B1,
\begin{equation}
\begin{array}{lll}
\bar{f}_{\rm loss}(t,t_{f1},t_{f2})& = &\frac{1}{|t_{f2}-t_{f1}|} 
\int_{t_{f1}}^{t_{f2}} f_{\rm loss}(t-t_f) dt_f  \\ 
\end{array}
\end{equation}
Aging can cause net stellar mass loss over time, for instance for
galaxies with no subhalo mass gain in a particular time step and a random
component $\Delta
M_{*,{\rm ran}}$ in Eq.~\ref{eq:dmstardmh} which is smaller than
stellar mass lost to aging.
\end{itemize}

Applying this algorithm to the histories of all the subhalos in the
simulation attaches to each galaxy (subhalo) a stellar mass and star
formation rate, at every output time.
As noted earlier, further observable quantities can be derived from these galaxy
stellar mass and
star formation rate histories.
The history of star formation in this prescription
could be combined with a stellar population synthesis model
(such as \citet{BC03}, \citet{Mar05} or FSPS \citep{ConGunWhi09,ConWhiGun10,ConGun10}), dust
model and initial mass function to give
luminosities in different bands, and other observational properties.
Morphological and gas properties are not as directly 
calculable.\footnote{The model also provides an estimate of the
  intra cluster light or ICL, studied in detail by \citet{Bec15}.  As the ICL is not as well characterized for
  the higher redshifts which are of most interest here, it is not 
considered.}  Galaxy positions, velocities, and their relation to the
cosmological cosmic web are also automatically available from
the dark matter simulation, although they will not be used below.

\section{Observational statistics}
\label{sec:simmeas}
Applying the \citet{Bec15} method as above to a simulation produces a
box of galaxies with stellar masses and star formation rates,
inhabiting host dark matter subhalos in their cosmological large scale
structure environment.  The full histories of the galaxies are also available and
some of their properties are discussed in \S \ref{sec:histories} below.  

Many variations of the model (see Table \ref{tab:dmsims}) were constructed.
For simplicity, only select examples are shown in the figures, while 
general trends are discussed
in the text.
The Bolshoi-P and TreePM runs have cosmological parameters closest
to the current best fit measurements, and
thus presumably the closest to the observations,  so the majority of 
results shown below are for bolp and tree.  
Although the Bolshoi run (bol) doesn't use the currently favored cosmological
parameters, the star formation efficiencies used in \citet{Bec15} and
here are based upon
it, and the cosmological parameters are similar to those used the \citet{Bec15}
simulation of the original model.  The bol and bolp differences give some indication of the
effect of varying cosmological parameters (however, the simulations also have different
time steps).  Results from the $M_{\rm
  vir}$ based models for the two Bolshoi simulations, bolmvir and
bolpmvir, are used to isolate effects of replacing the smoothed and
constrained $SAM\_M_{\rm vir}$ masses with $M_{\rm vir}$ and
calculating accreted mass using Eq.~\ref{eq:nosam} instead.
\citet{Bec15} used  the Rockstar 
\citep{rockstar} $M_{\rm vir}$ and
Eq.~\ref{eq:samacmass} to calculate accreted mass; this combination
is also used for the bolpmmp and bolmmp runs.
These cases give 
similar results to bolmvir and bolpmvir, respectively, aside from
slightly larger numbers of low stellar mass galaxies, slightly more
star forming galaxies at most stellar masses, and somewhat lower
stellar
masses for a given halo mass. 
They will not be discussed further.

Below, the stellar mass-star formation rate
($M^*-SFR$) relations for the simulations and the (related) specific
star formation rates are shown, along with the 
separation between star forming and quiescent galaxies found by
\citet{Mou13} for PRIMUS.  For each sample, this diagram is used to estimate the split between
star forming and quiescent populations, and then stellar mass
functions are calculated for star forming, quiescent and all galaxies
in \S\ref{sec:smfcomp}, and compared to observations.
Last, but not least, the
simulated stellar mass to halo mass relations are compared to two
relations found from observations by \citet{BWCz8} and \citet{MosNaaWhi13}.
Quantities are shown for (sometimes some subset of) redshifts $\sim$0, 0.57,
0.91 and $\sim 2$.
\begin{figure*}
\begin{center}
\resizebox{3.3in}{!}{\includegraphics[clip=true]{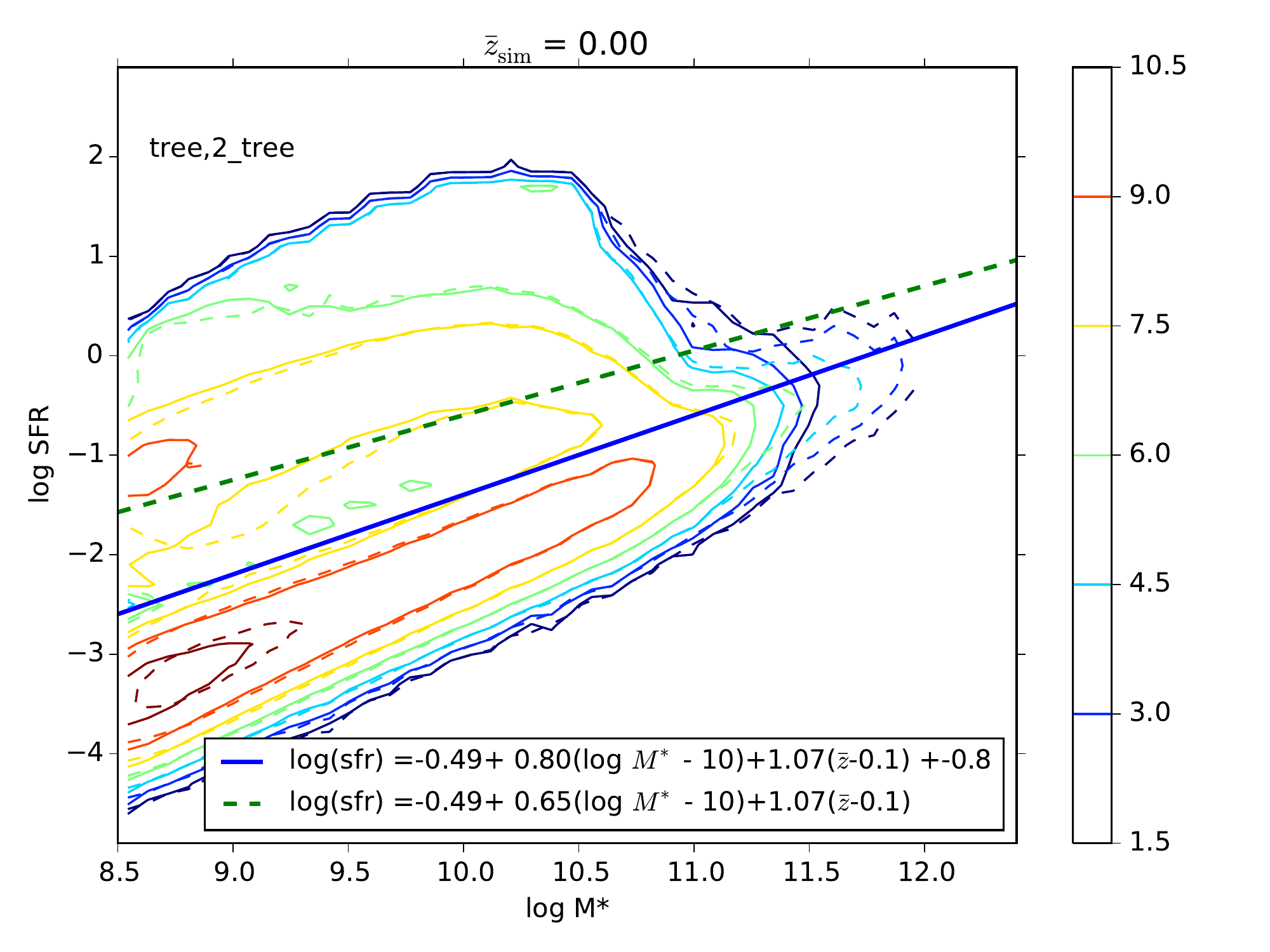}}
\resizebox{3.3in}{!}{\includegraphics[clip=true]{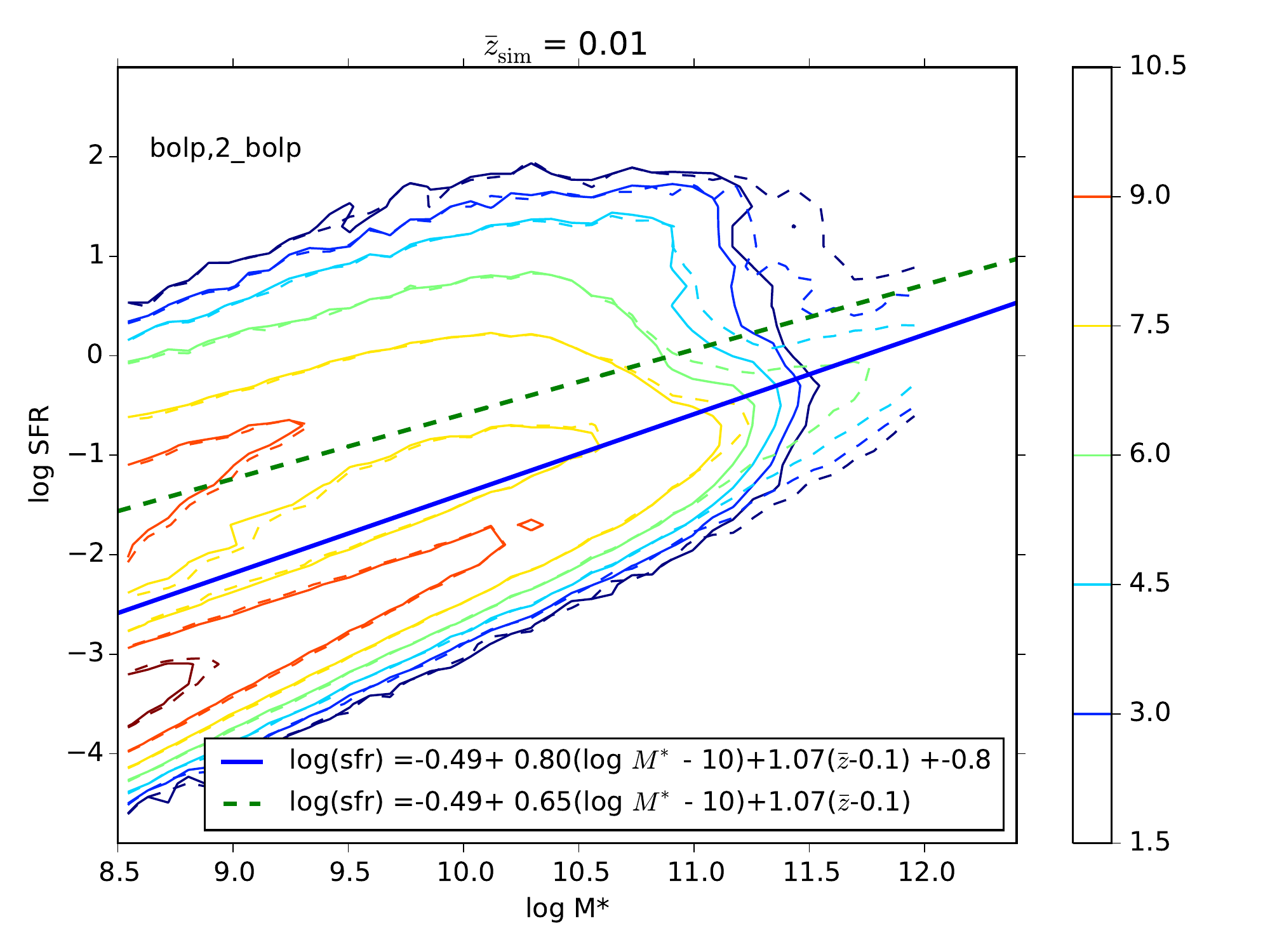}}
\resizebox{3.3in}{!}{\includegraphics[clip=true]{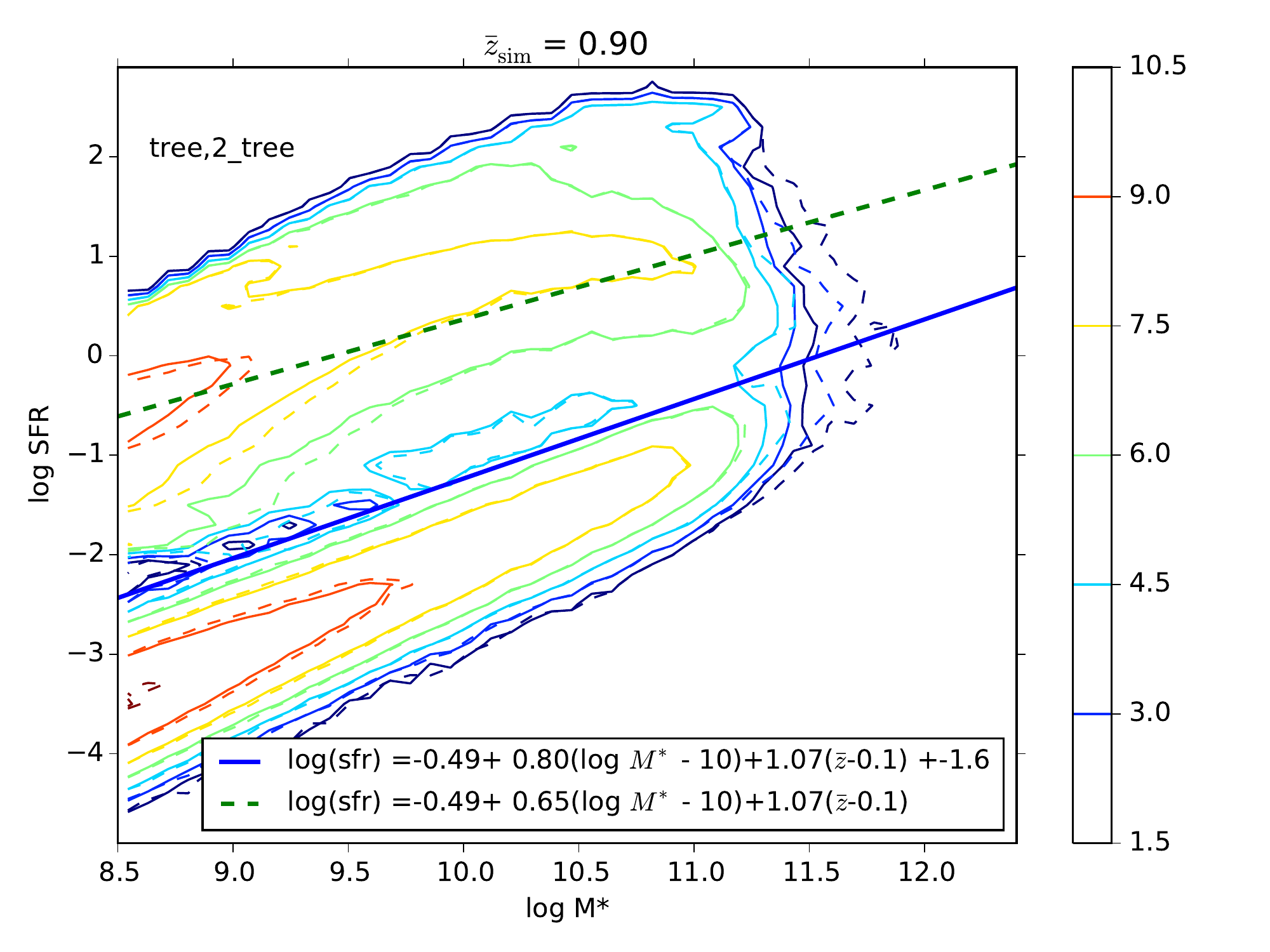}}
\resizebox{3.3in}{!}{\includegraphics[clip=true]{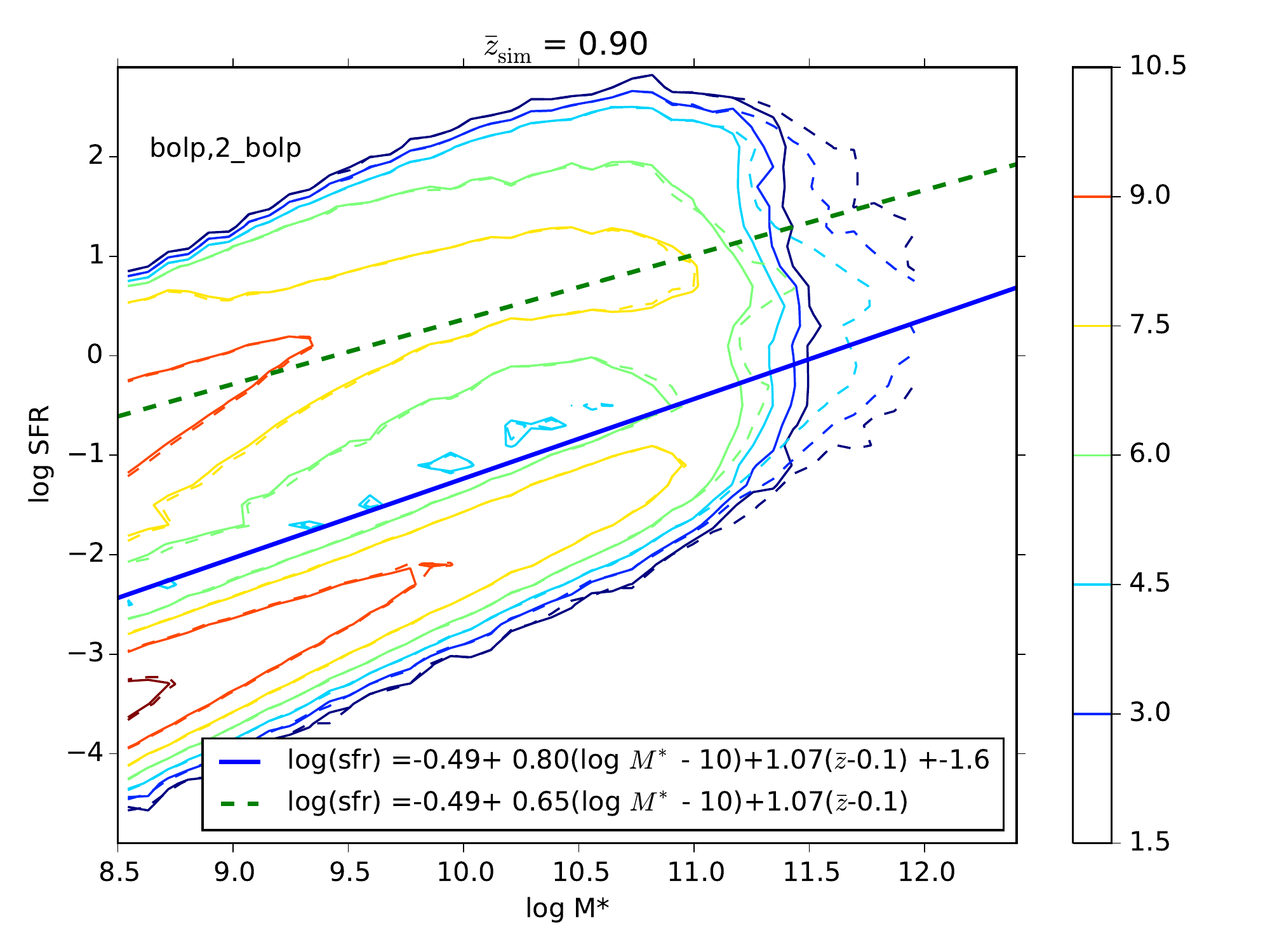}}
\resizebox{3.3in}{!}{\includegraphics[clip=true]{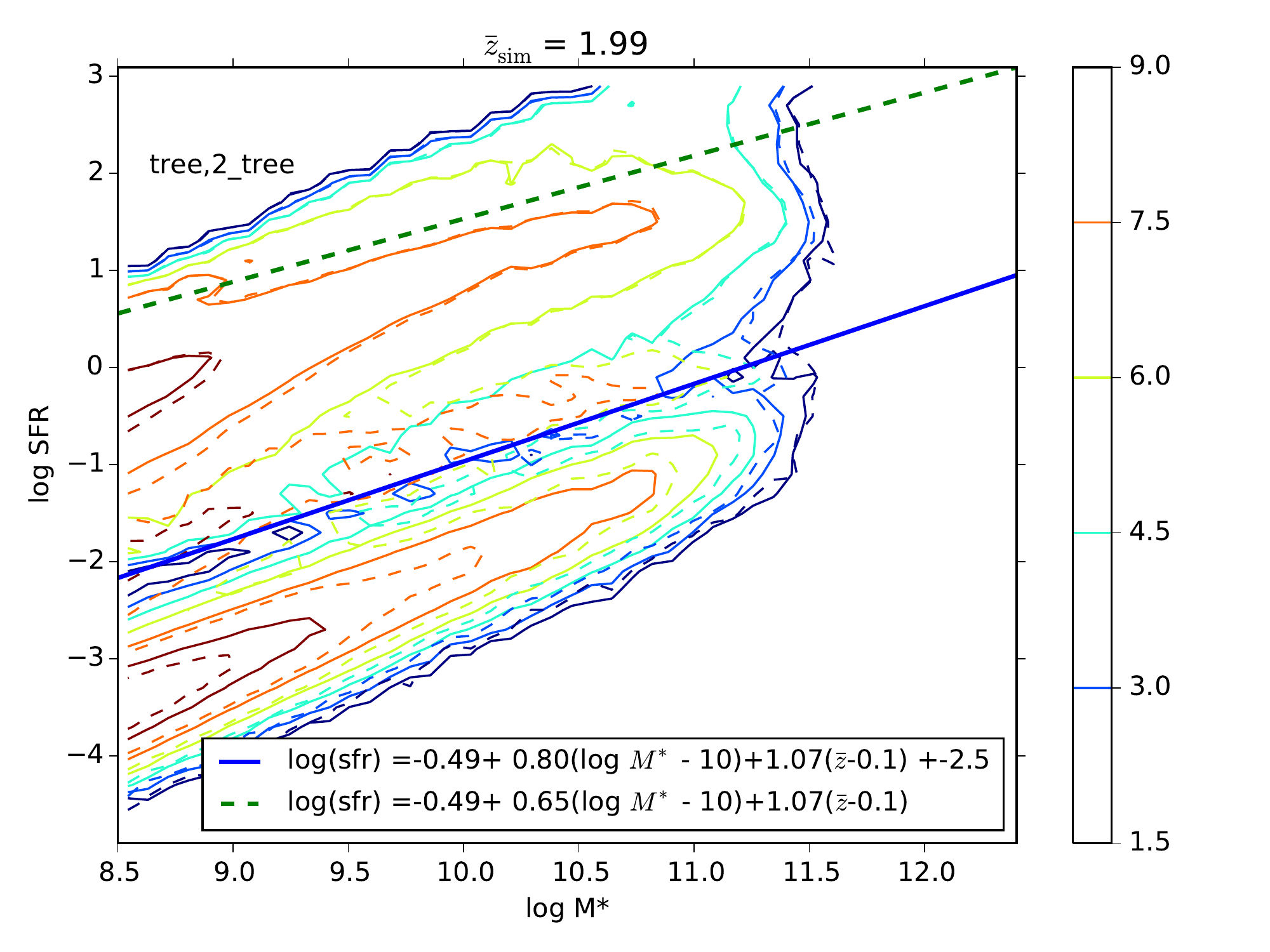}}
\resizebox{3.3in}{!}{\includegraphics[clip=true]{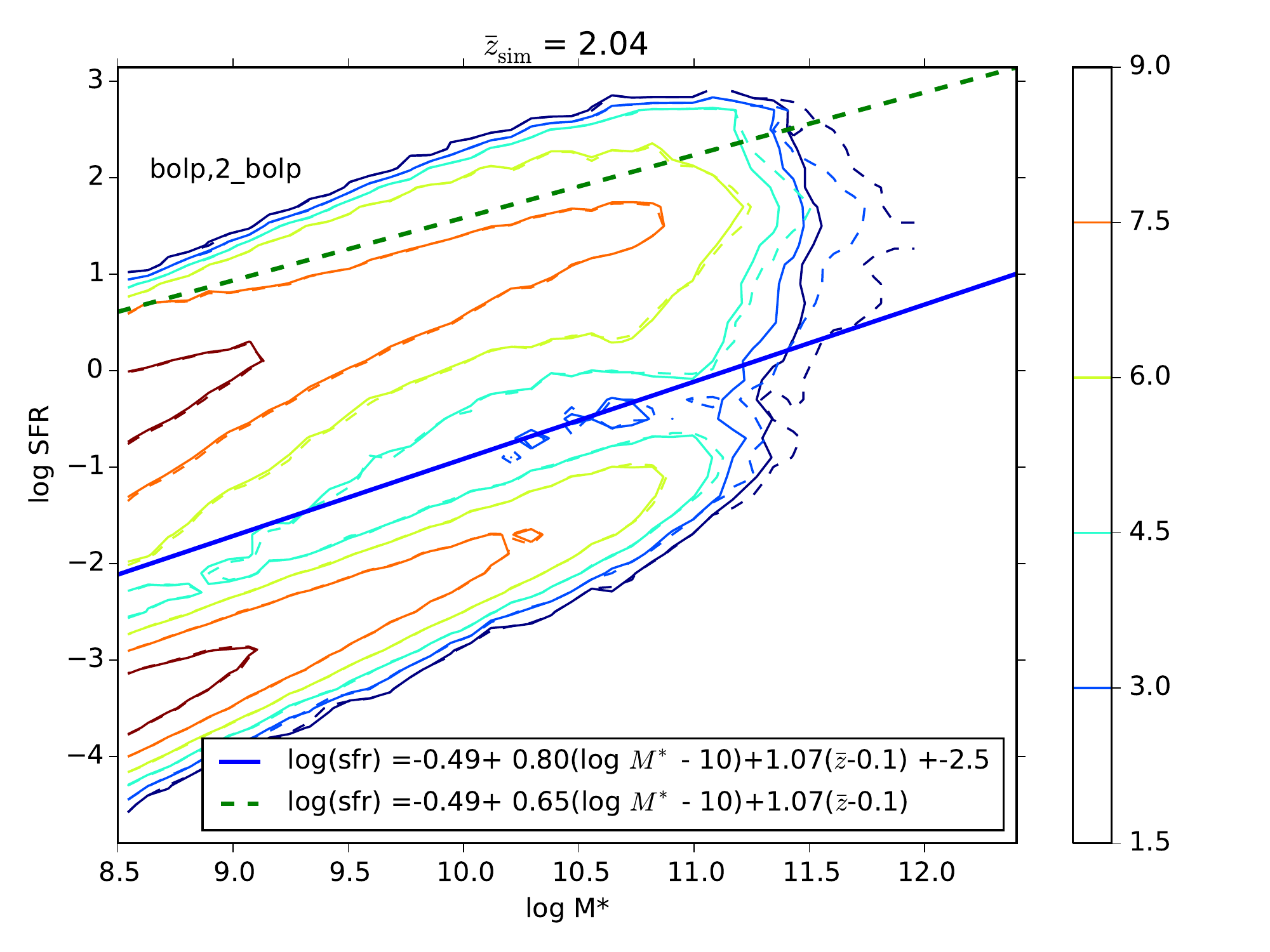}}
\end{center}
\caption{Star formation rate as a function of stellar mass, for 1
  progenitor models (solid lines) and 2 progenitor models (dashed
  lines), for tree, 2tree (left) and bolp, 2bolp (right).
The slope and offset for the line separating the models are as listed
and were chosen by eye.
The dashed line, Eq.~\ref{eq:primuscut}, is from \citet{Mou13} and
does not separate the two peaks.
The difference between the \citet{Mou13} sample and the models may be
in
part because the former calculate
stellar mass using FSPS
\citep{ConGunWhi09,ConWhiGun10,ConGun10}, while the star formation
efficiencies for the models constructed here use BC03
\citep{BC03} stellar masses.
Contours are given by
arcsinh of galaxy counts in each pixel, to enhance the range shown.
}
\label{fig:mstarsfr} 
\end{figure*}
\begin{figure*}
\begin{center}
\resizebox{3.2in}{!}{\includegraphics[clip=true]{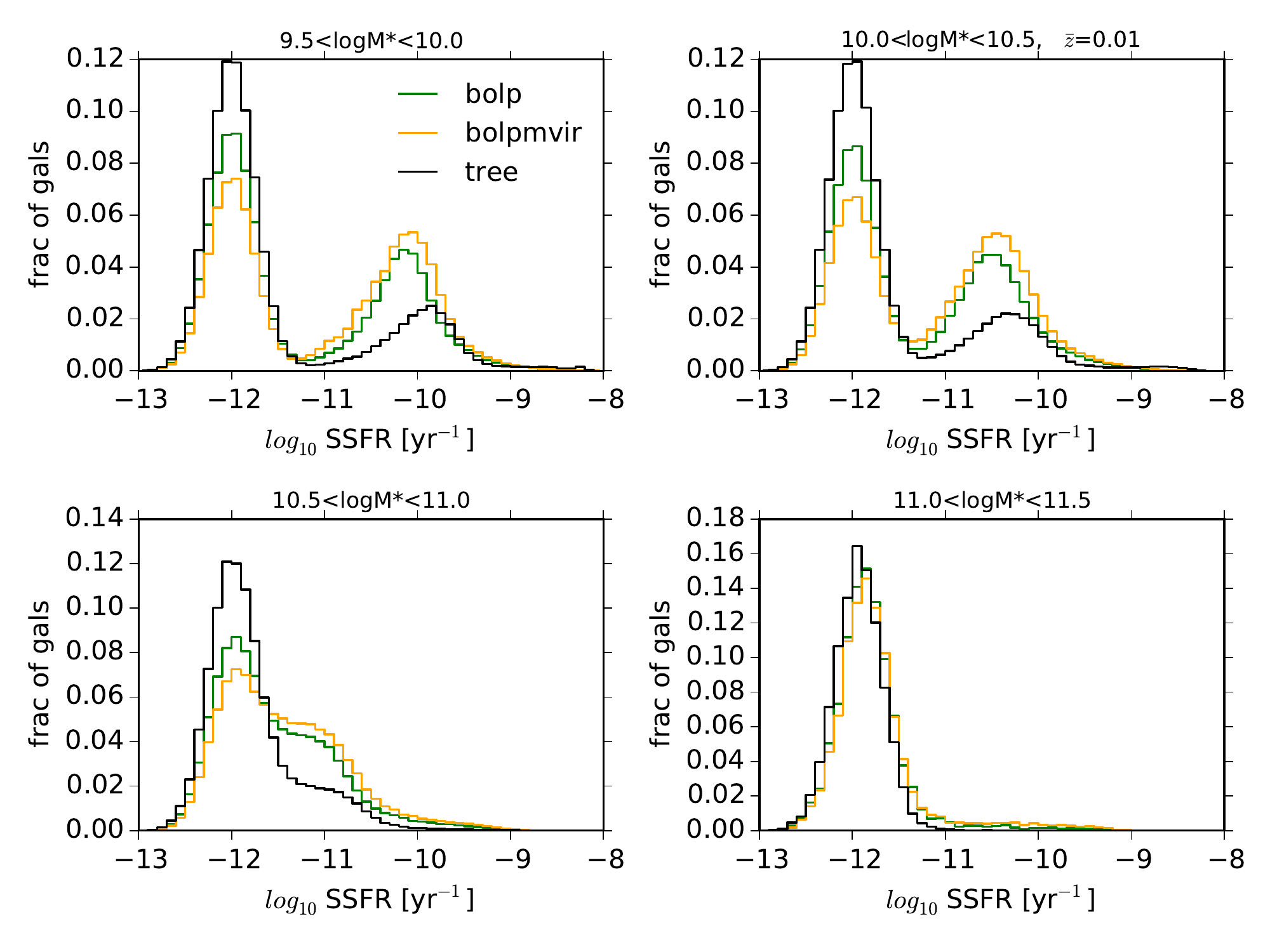}}
\resizebox{3.2in}{!}{\includegraphics[clip=true]{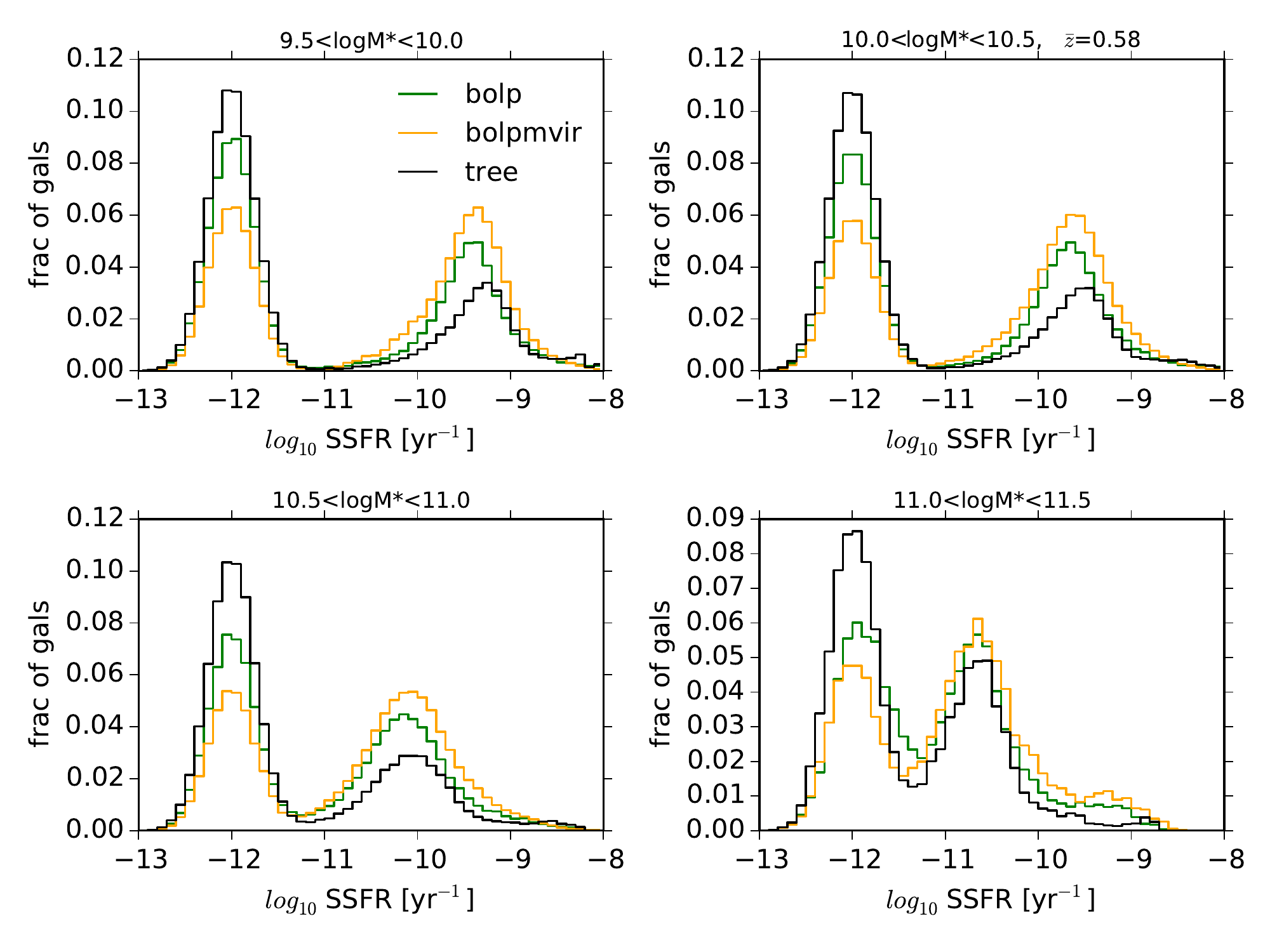}}
\resizebox{3.2in}{!}{\includegraphics[clip=true]{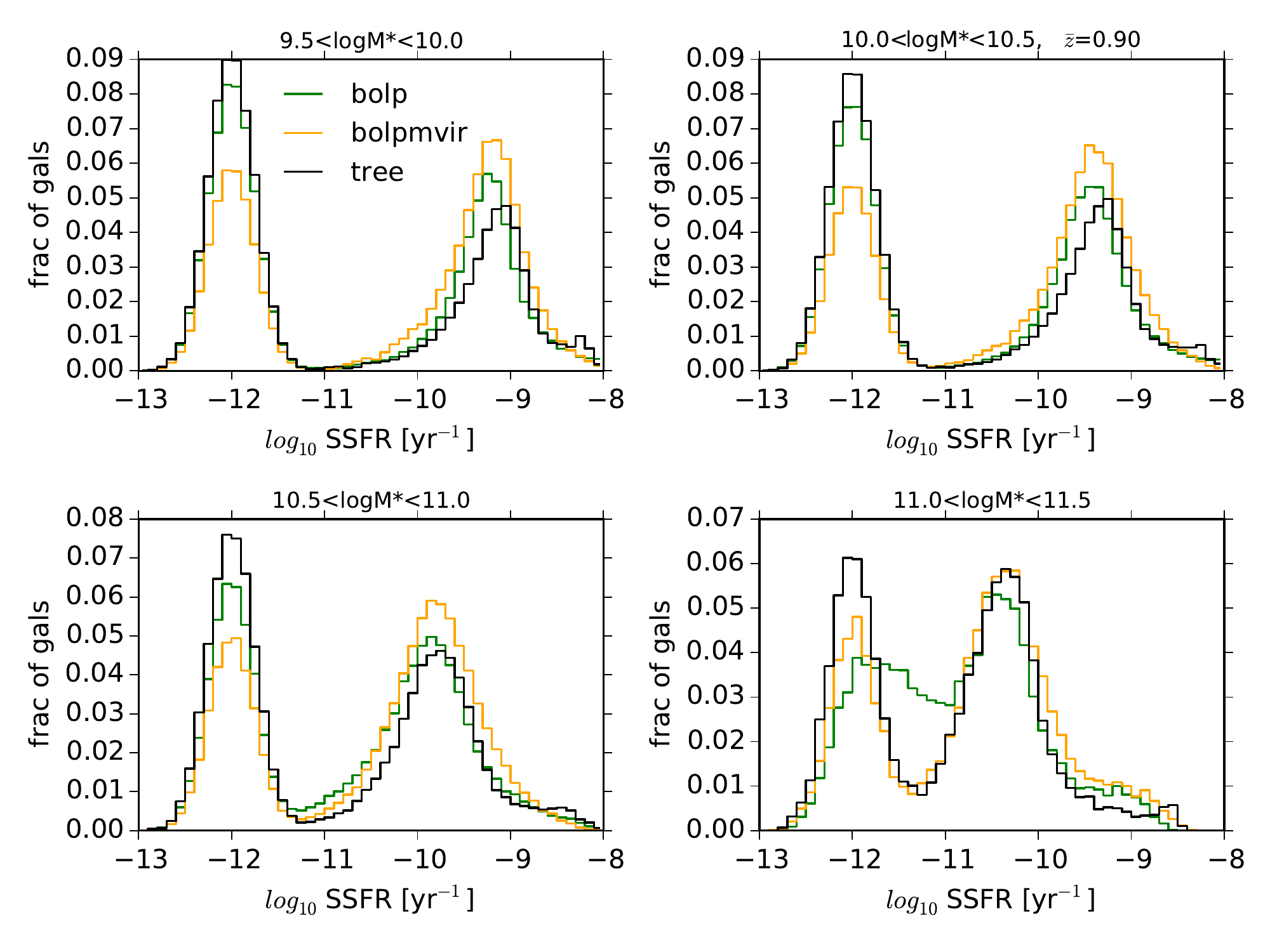}}
\resizebox{3.2in}{!}{\includegraphics[clip=true]{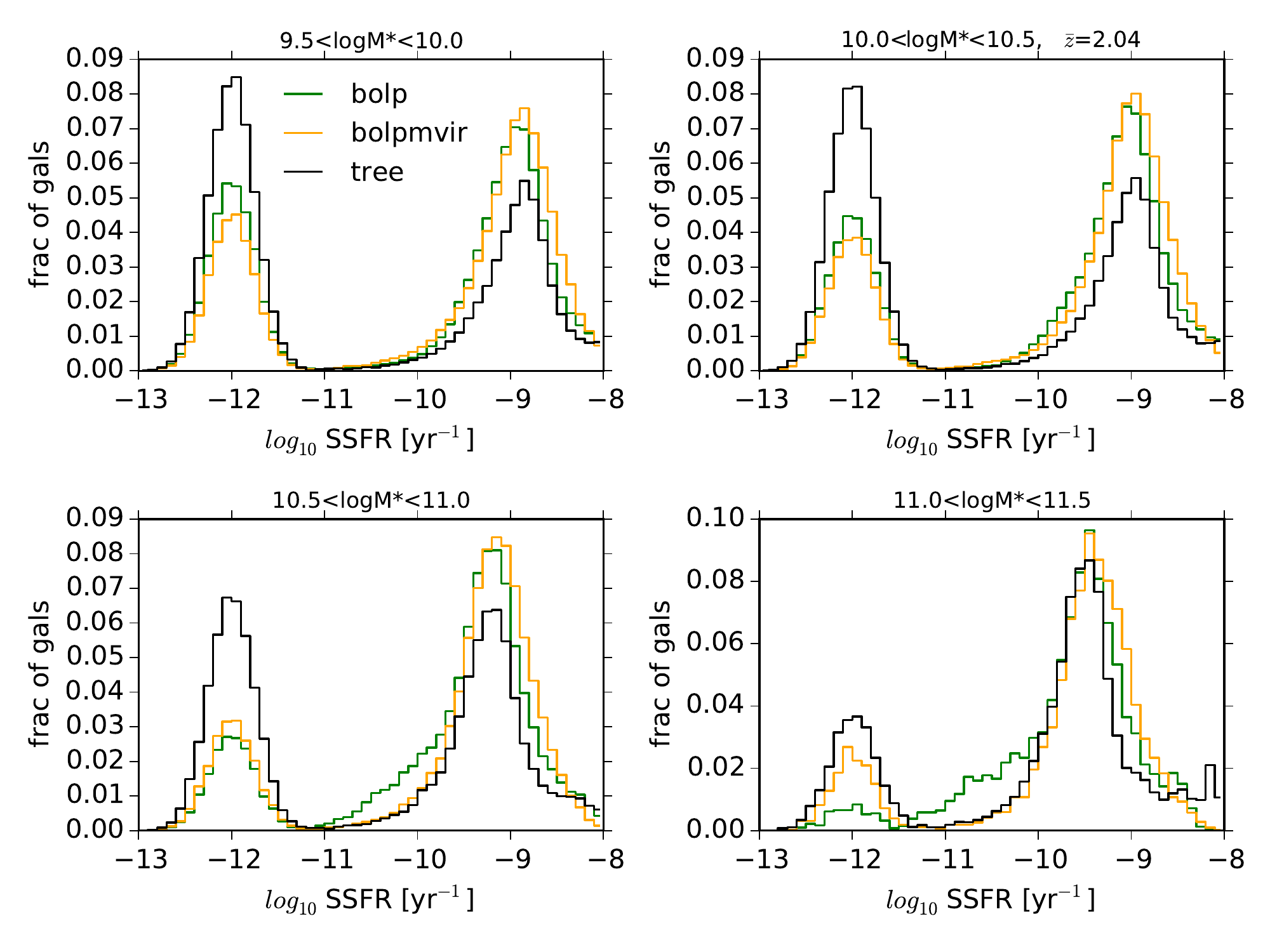}}
\end{center}
\caption{Specific star formation rate in four stellar mass bins (as
  indicated in each panel) for the tree, bolp and bolpmvir models at
  redshifts 0.01, 0.58, 0.9 and $\sim 2$.  The 1 and 2 progenitor
  models are very similar in specific star formation rates, so only the 1 progenitor models
  are shown.  The left side peak is highest for the tree model,
  because the tree model satellites, aside from mergers, have only
the random star formation term, Eq.~\ref{eq:dmstardmh}, as their
star formation rate.}
\label{fig:ssfr} 
\end{figure*} 

\subsection{Stellar mass vs. star formation rate}
\label{sec:mstarsfr}
To start,
the stellar mass-star formation rate is shown in
Fig.~\ref{fig:mstarsfr}, for the TreePM based tree, 2tree models
(left,
solid and dashed respectively) and the Bolshoi-P based bolp,
2bolp models (right, solid and dashed respectively), for a sample of
redshifts.  The bol model is similar to bolp and the bolmvir
model is similar to tree, in that the latter two (which use
instantaneous rather than $SAM\_M_{\rm vir}$ mass) have a slightly
less smooth boundary between different regions (for example at
redshift $\sim 2$).
For all the model implementations, there are two peaks clearly visible.
The dashed green line is the separation found by
\citet{Mou13} for PRIMUS (see their Figure 1).  
In that case \footnote{Note the typo in \citet{Mou13}
Eq. 2 which is corrected in-line in the text directly above.} 
\begin{equation}
\log (SFR_{\rm min})= - 0.49+0.65 (\log M^* -10)+1.07 (z - 0.1)
\label{eq:primuscut}
\end{equation}
The by-eye separation between star forming and quiescent galaxies
is shown as a solid blue line.  It lies at lower star formation rate,
and with a steeper slope, than the dashed green PRIMUS separating line.
For the bol and bolmvir models the
dividing line also has slope 0.8 (rather than 0.65 as in 
Eq.~\ref{eq:primuscut} above), but shifts instead by 
(-0.5,-1.3,-1.5,-2.8) at redshifts $\sim (0, 0.6,0.9,2)$  (compared to (-0.8,-1.3,-1.6,-2.5) for the models
above).
The observed separation
between star forming and quiescent galaxies
moves to higher star formation rates as redshift increases (this is
seen also in other observations, e.g. \citet{Tas15}).  This trend to
higher star formation rate at higher redshift is
weaker
in the simulated samples.

Part of the differences may be due to the
different
stellar mass definitions used (the stellar
mass assumptions are given for the the observational dashed separating
line (PRIMUS), and the star formation efficiencies from \citet{BWCz8} (bwc\_comp) in Table \ref{tab:smfobs}; a comparison of
many different observational star formation rates as a
function of $M^*$ is found in \citet{Spe14}).
Another difference is that the center of the quiescent branch is at a
fixed position (by hand) at
all redshifts in the simulations,
while observationally the quiescent branch moves to higher star formation rates at
higher redshifts.  This fixing of the quiescent branch also pins the
 star forming/quiescent dividing line to lower star formation rates.
Presumably the quiescent branch should evolve in position (and perhaps scatter)
with redshift.

\begin{table*}
\begin{tabular}{l|l|c|c|c|c|c|c|} 
Name&Reference&$z$ bin edges&area & SPS model
& Dust&IMF & Active vs.\\
&&& & 
& & & Quiescent\\
\hline
SDSS-GALEX&\citet{Mou13}&$\bar{z}\sim$0.1&6956 deg${}^2$&FSPS
&\citet{ChaFal00} &Chabrier &SFR-$M^*$
\\
& Table 3& & & & & &Eq.~\ref{eq:primuscut}   \\
\hline
PRIMUS\_fsps&\citet{Mou13}&(0.2,0.3,0.4,&5.5 deg${}^2$ 
&FSPS&\citet{ChaFal00}&Chabrier&SFR-$M^*$ \\
& Table 4&0.5,0.65,0.8,1.0) & & & & &Eq.~\ref{eq:primuscut}  \\
PRIMUS\_bc03&www.peterbehroozi.com&as above&as above
      &BC03&\cite{BlaRow07}&Chabrier&n/a \\
\hline
ZFOURGE&\citet{Tom14}&(0.2,0.5,0.75,1)&316 arcmin${}^2$&BC03&FAST \citep{KrivanLab09}
& Chabrier&UVJ\\
& Table 1& & & & $A_V \in$ [0,4],$Z_\odot$& &   \\
  \hline
COSMOS/ &\citet{Muz13}&(0.2,0.5,1.0,&1.6
                                                   deg${}^2$&BC03&\citet{Cal00}&Kroupa&UVJ
  \\
Ultravista& cosmos2.phy.tufts.edu/  & 1.5,2.0,2.5,& & & &(converted) &   \\
&$\sim$danilo/Downloads.html&3.0,4.0)&&&&&\\
\hline
VIPERS&\citet{Mou16}&(0.2,0.5,0.8& $>$22 deg${}^2$&
                                                BC03&\citet{Cal00}&Chabrier&NUVrK
  \\
& Table 2 & 1.1,1.5)& & &+ 2 other & & \\
\hline
\hline
\hline
\end{tabular}

\begin{tabular}{l|l|c|c|c|c|c|c|c|} 
Compilation&Reference&$\bar{z}$& SPS model
& Dust&IMF & Active vs.\\
&&
& & & & Quiescent\\
\hline
bwc\_comp& \citet{BWCz8}&(0.,0.5,1.,2.)&
                                   BC03&\citet{BlaRow07}&Chabrier&
                                                                           N/A
  \\
& Fig. 3 & & & & & \\
\hline
Hen15&\citet{Hen15}&(0.1,0.4,1.,& 
                                                         \citet{Mar05}&&Chabrier&
  \\
&
http://galformod.mpa-garching.mpg.de/&2.,3.) &(converted) & & &\\
& public/LGalaxies/figures\_and\_data.php & & & & & \\
  \hline
\end{tabular} %
\caption{Observational data sets included.  Additional references are \citet{Str16} for ZFOURGE and \citet{Mar09,Mar10} for COSMOS/Ultravista.  The conversions used for Kroupa
  to Chabrier IMF and for \citet{Mar05} to BC03 stellar population
  synthesis are described in the text. }
\label{tab:smfobs} 
\end{table*}

Slices of specific star formation rate in different mass bins can also be
calculated, these are shown in Fig.~\ref{fig:ssfr} for the
tree, bolp and bolpmvir models (only
the 1 progenitor models are shown here, there is not much difference
for the 2 progenitor models).  The tree model has the most quiescent
galaxies, in part because the tree satellites are almost all quiescent (as
their subhalo masses are fixed to their infall subhalo mass unless they merge).  The
satellites in the bolp and bolpmvir versions can
(and do) have subhalo and thus stellar mass gain.  The bolpmvir model,
with 
instantaneous mass gain, has more
galaxies with high specific star formation rates than the smoothed and
constrained $SAM\_M_{\rm vir}$ bolp model (also seen for bolmvir
compared to bol). The instantaneous
mass $M_{\rm vir}$ is presumably more stochastic in mass gain than its
smoothed and constrained counterpart.
At the highest stellar masses, the smoothed and
constrained bolp specific star formation rates often show a less clean
bimodality than the bolpmvir or tree specific star formation rates.
\subsection{Stellar mass functions}
\label{sec:smfcomp}
The relations used to get stellar mass
and star formation rates were measured for
average binned halo mass and stellar mass gains.
The model here uses these rates to
assign stellar mass and star formation rates to individual galaxies.
As the average of a non-linear function of some quantity is not the same as the
non-linear relation on the average of that quantity, the resulting
average stellar mass
functions and star formation rates may or may not agree with
observations.
Here, the resulting stellar mass functions are compared to each other
and observations
for all, quiescent and star forming galaxies.

\subsection{Observational stellar mass functions}
\label{sec:obssmfs}
Before comparing with observations, it should be noted that the
available observational stellar mass functions
do not all completely overlap, even including their error
bars.
In Table \ref{tab:smfobs}, the different stellar mass function
observations used are listed, along with many of their properties.  At
the bottom are two compilations, from \citet{BWCz8} and \citet{Hen15}.
Although stellar mass is motivated
in part by the wish to reduce differences between observations (such
as different wavebands),
assumptions made in calculating stellar masses can differ between
observations and between the models used to predict the stellar masses.  The surveys
whose data are shown below 
were analyzed with a variety of stellar population synthesis models,
dust models, initial mass functions (IMF), and cosmological
parameters, as listed in Table \ref{tab:smfobs}. 

The \citet{Bec15} model for assigning stellar mass and star formation rates
has specific choices for these properties built in as well.  It uses star formation efficiencies
from \citet{BWCz8}, corresponding to a \citet{BC03}(BC03) stellar
population synthesis model, with \citet{BlaRow07} dust and a Chabrier
IMF (they convert all measurements they use to this common basis,
i.e. that used by the bwc\_comp compilation in Table \ref{tab:smfobs}).

For some of the observational data available at these redshifts,
the stellar population synthesis models are instead either FSPS 
\citep{ConGunWhi09,ConWhiGun10,ConGun10} or \citet{Mar05}.  \citet{Mar05} is used
by the \citet{Hen15}; the conversion to BC03 is taken from that paper,
i.e. 
\citep{Dom11}, $\log_{10}M^*_{\rm Mar} = \log_{10}M^*_{\rm BC03} -
0.14$.
Conversions between BC03 and FSPS are not available.
Dust models also include those by \citet{ChaFal00,Cal00} and others, 
and the COSMOS-Ultravista \citep{Muz13} observations use 
a Kroupa IMF instead of a Chabrier IMF.  The stellar mass
functions based on the Kroupa IMF are converted
by rescaling the Kroupa $M^*$ by 0.61/0.66 \citep{MadDic14}. 

In addition to variations in stellar mass definitions, different surveys have
different definitions of quiescent and star forming (shown in the last
column of Table
\ref{tab:smfobs}). 
\citet{Mou13} use
the SFR-$M^*$ relation in
Eq.~\ref{eq:primuscut} above, also shown in Fig.~\ref{fig:mstarsfr} as
a dashed line in each panel.  It is not a good fit for the models here
(again because the center of the quiescent branch in the models does not change
with redshift,  and likely because also PRIMUS uses a different population synthesis model and
dust model than that used to calculate the star formation efficiencies
in \citet{BWCz8}).  In more detail,
the PRIMUS SFR is calculated via iSEDfit (as is $M^*$), which takes
measurements in 
all of their specific filters.  
In contrast, \citet{Tom14,Muz13} separate quiescent 
and active stellar galaxies
 using UVJ and \citet{Mou16} separate colors using NUVrK. 
For comparison with the simulations, the stellar mass functions for the quiescent and star forming
populations 
are simply plotted as given
(after any known conversions for stellar mass or IMF have been included as
above).
As can be seen below, sometimes, but not always, the UVJ and NUVrK separated stellar mass functions
have strong overlap with the stellar mass functions found using
the star formation -$M^*$ cut, Eq.~\ref{eq:primuscut} of \citet{Mou13}.

Conversions between these different definitions of stellar mass and
quiescence are not always
straightforward without reanalyzing the full observational sample, and
the data required for conversions are not always available.
Comparisons of one fixed observational data set with several different stellar mass
modeling assumptions are found in the appendix of \citet{Mou13} for
PRIMUS; tools for such comparisons such as EZGAL \citep{ManGon12} can
also be used.

One additional consideration is that
 the observed volumes are often still relatively small, although
efforts have been made to estimate the sample variance errors (for
example, the sample variance estimate in ZFOURGE follows
\citet{Mos11}).  It is also 
possible that the covariance of the scatter between different stellar mass bins
described in
\citet{Smi12} may need to be included at some point.

\subsection{Comparison between observations and the models}
The above observational stellar mass functions were compared to the 
different models constructed here, at a range of redshifts.

\begin{figure*}
\begin{center}
\resizebox{3.4in}{!}{\includegraphics[clip=true]{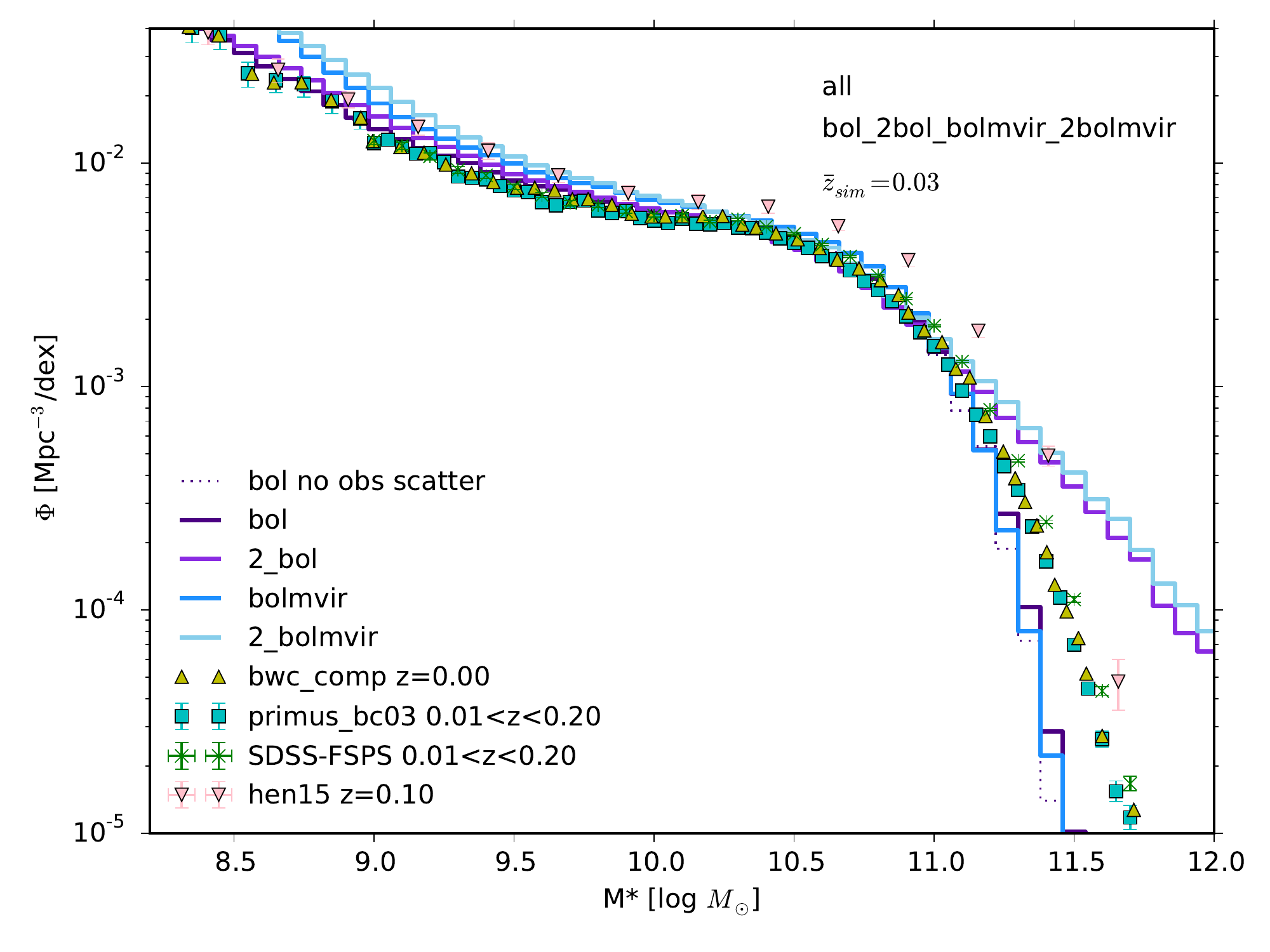}}
\resizebox{3.4in}{!}{\includegraphics[clip=true]{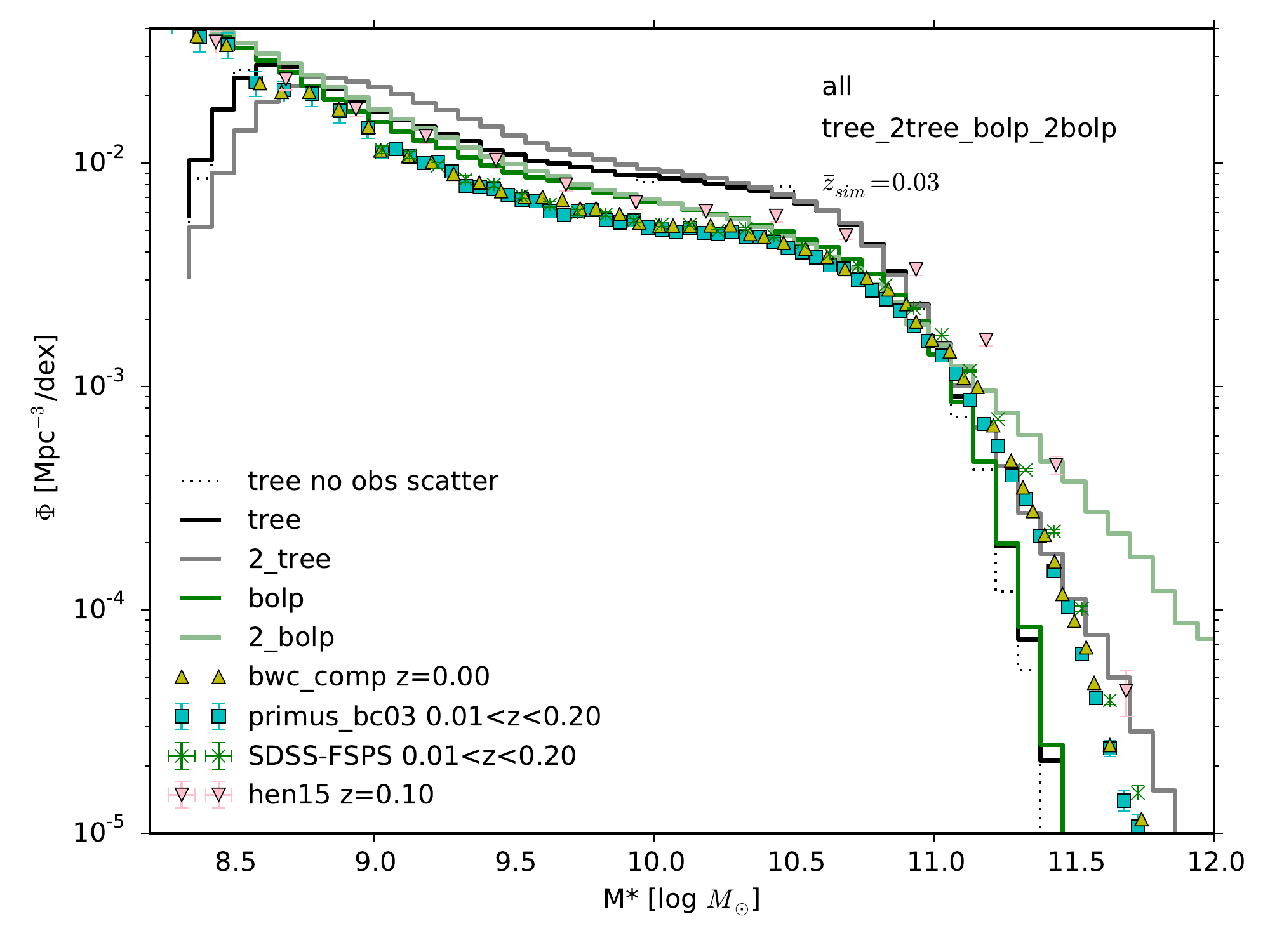}}
\end{center}
\caption{Stellar mass functions for bol and bolmvir models (left)
  and for tree and bolp models (right), for
  all galaxies at redshift $\sim 0$.  Histograms are for the
  simulations, while points and smooth lines are observations,
  described in Table \ref{tab:smfobs}.  The heavier histograms are for
  the 1 progenitor variants, and all solid histograms
include an observational scatter in $M^*$ (see text).
The dotted
histogram is the data before including scatter, shown for a single model in each panel.
The best fit to observations at low stellar mass is for the bol and 2bol models,
which
coincide with the cosmology and simulation used to calculate the star
formation
efficiencies implemented in all the models.
Changing the cosmology ((2)bolp, at right) increases the
stellar mass function at low stellar mass, as does using the
instantaneous $M_{\rm vir}$ ((2)bolmvir) or instantaneous $M_{\rm
  fof}$ ((2) tree, which also has the same cosmology as bolp).
At high stellar mass, the observations tend to fall between the 1 and
2 progenitor variants (except for 2tree), implying some combination
may work in general.
}
\label{fig:smfz0}
\end{figure*}

A first example is shown in Fig.~\ref{fig:smfz0}, for the bol,
bolmvir, bolp and tree models at $z=0$, along with observational data from
Table \ref{tab:smfobs} at the same redshift.  The histograms are from the simulations,
and include observational scatter via
the lognormal ($\log_{10}M^*$) distribution of \citet{BWCz8},
with variance $\sigma=0.07 + 0.04 z$ as a function of redshift.
The dotted histogram shows one of the models (as indicated) without this observational
scatter.  
In this and all figures following, the points and smooth lines are various observational data
sets, as indicated and further detailed in Table \ref{tab:smfobs}.
(The smooth lines are Schechter fits provided as part of the analyses the observations.)

The bol and bolmvir models
are based upon the 
simulations (with a disfavored 
cosmology) used to calculate the original star formation efficiencies
($SFE(M_h)$ in Eq.~\ref{eq:dmstardmh}, by
\citet{BWCz8}).  These may be expected to have the best match to
observations as a result.  The difference between bol and bolmvir is the choice of mass and method for
calculating accreted mass (see Table \ref{tab:dmsims}). 
For both panels at $z=0$, the models based upon the $SAM\_M_{\rm
  vir}$ smoothed and constrained halos masses give lower stellar mass
functions at low stellar mass than the models based upon instantaneous
mass ($M_{\rm vir}$ for bolmvir and $M_{\rm fof}$ for tree).  The
lowest stellar mass function at low stellar mass is for the bol models, which has the best
agreement with observations.  At high stellar mass, except for the
tree variants, the observations fall between the 1 and 2 progenitor
models, implying some criteria to mix the two will work (such as the
simulation specific criterion of \citet{Bec15}).
\begin{figure*}
\begin{center}
\resizebox{3.4in}{!}{\includegraphics[clip=true]{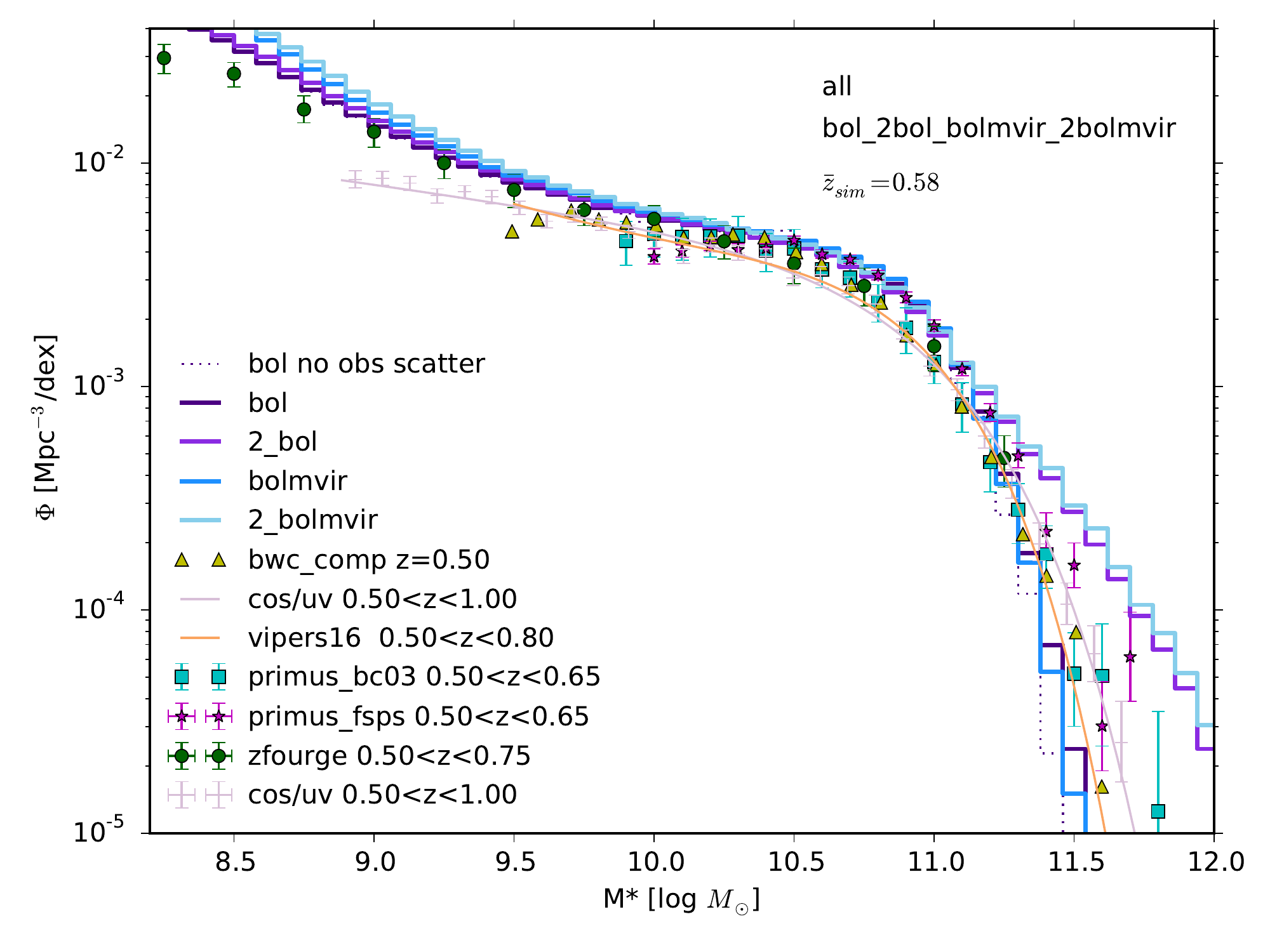}}
\resizebox{3.4in}{!}{\includegraphics[clip=true]{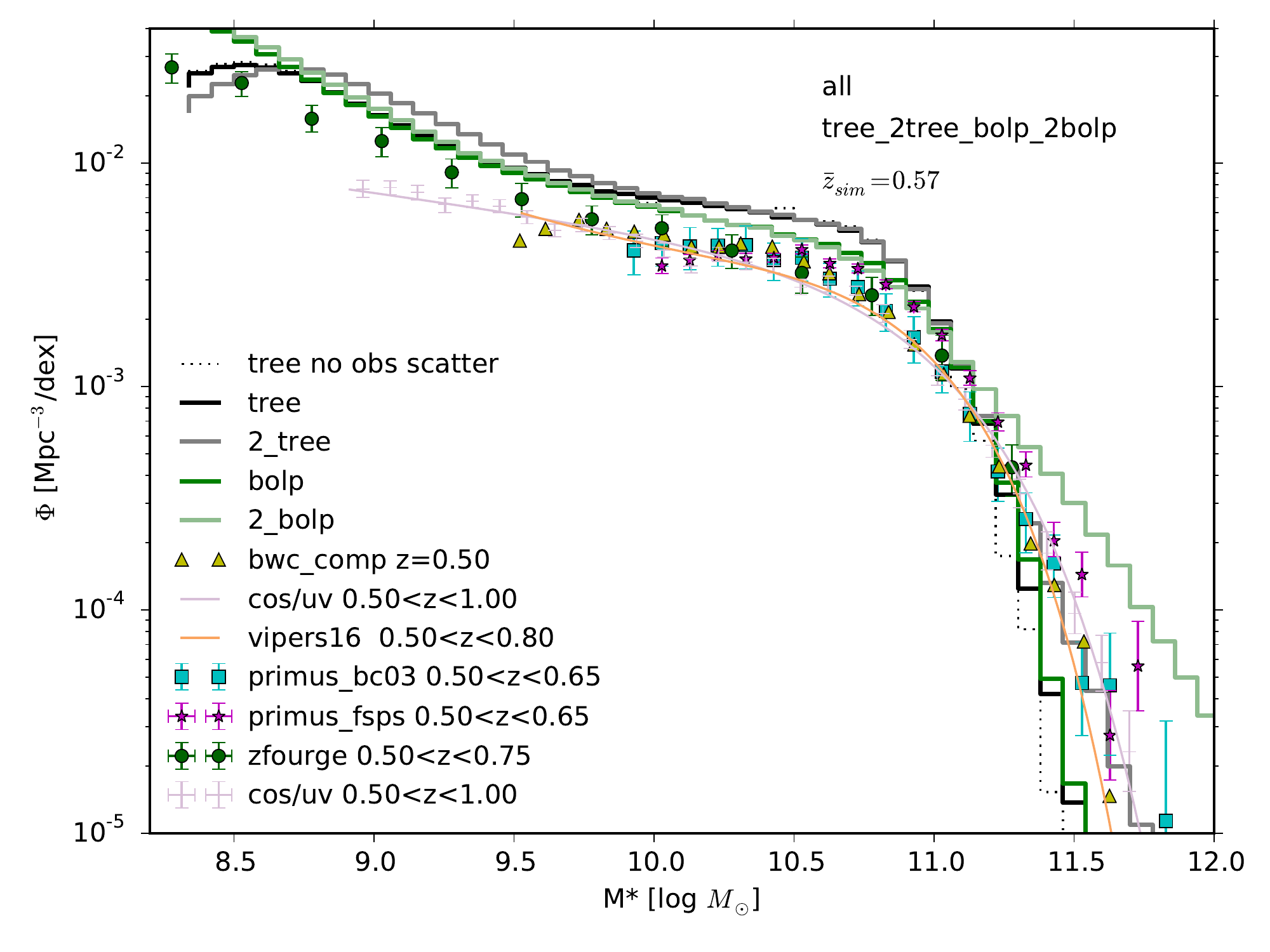}}
\resizebox{3.4in}{!}{\includegraphics[clip=true]{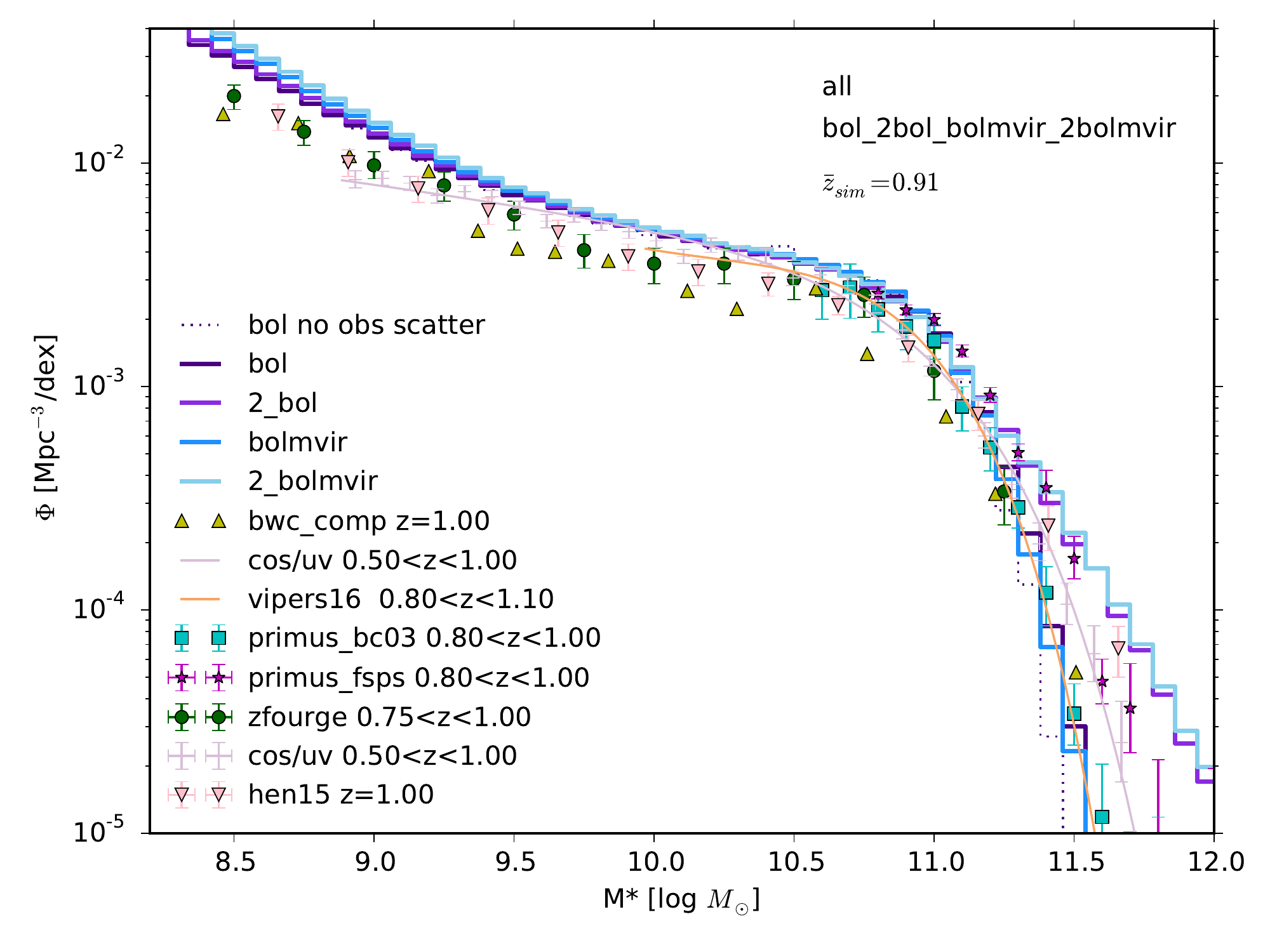}}
\resizebox{3.4in}{!}{\includegraphics[clip=true]{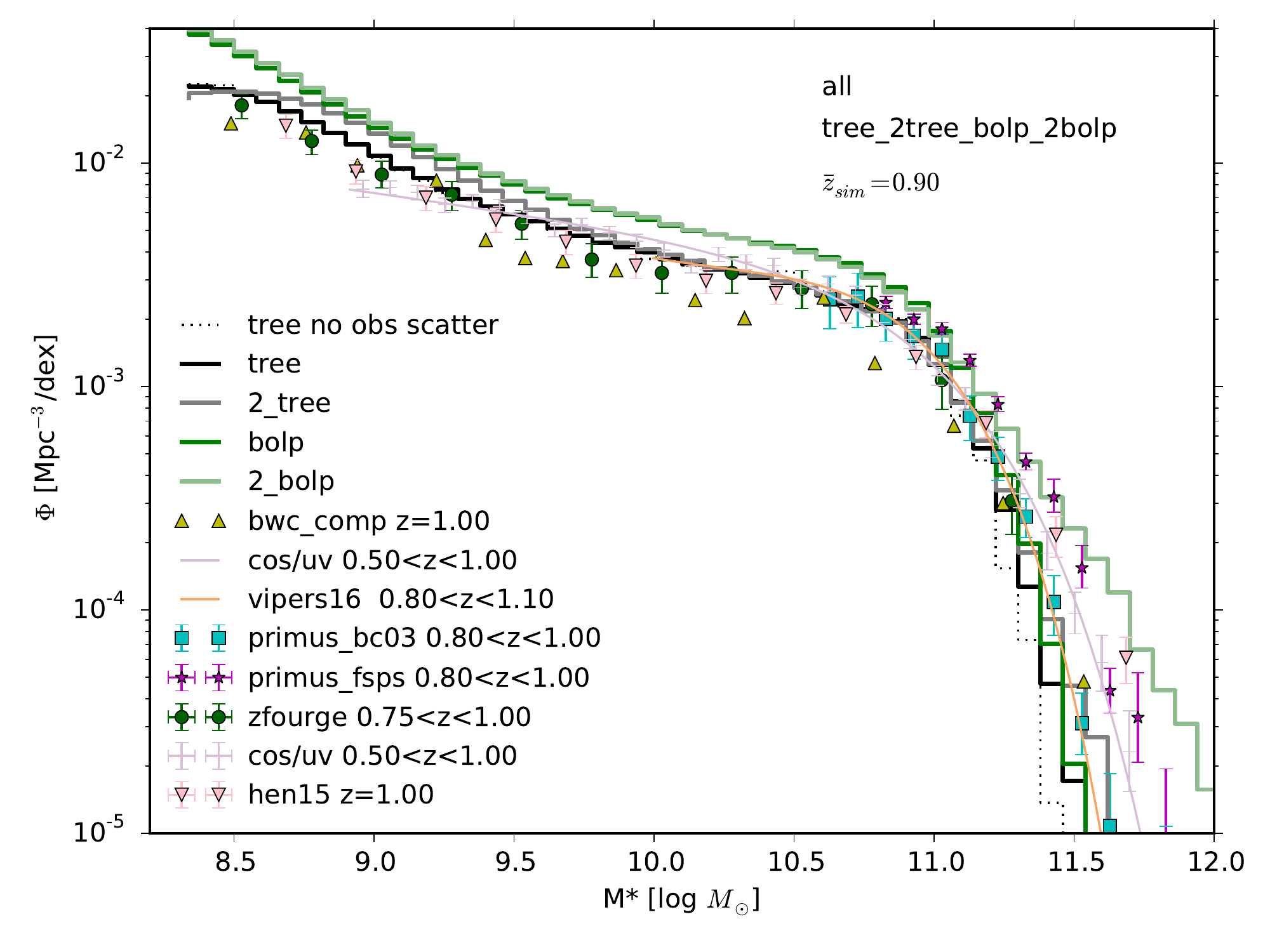}}
\resizebox{3.4in}{!}{\includegraphics[clip=true]{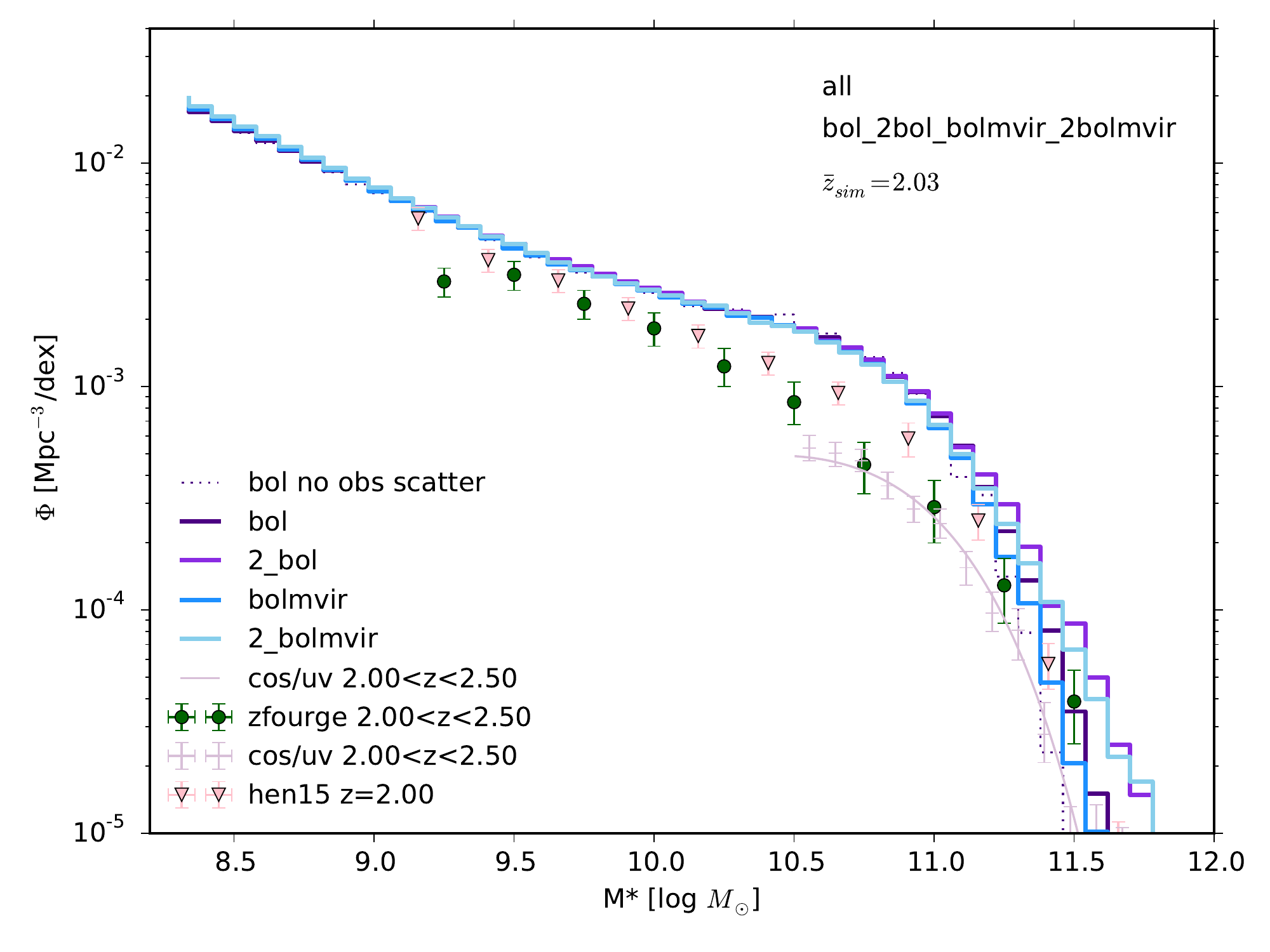}}
\resizebox{3.4in}{!}{\includegraphics[clip=true]{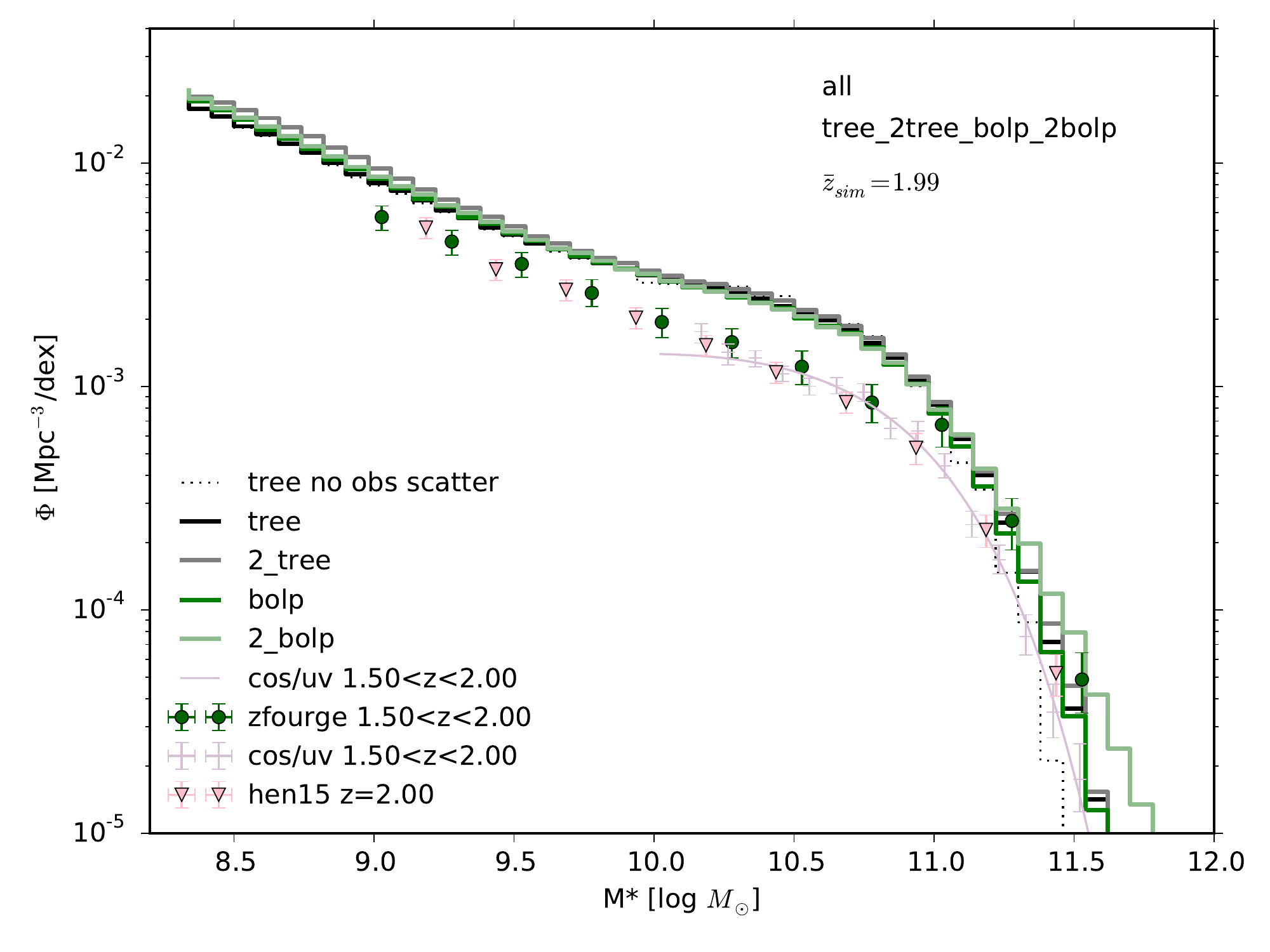}}
\end{center}
\caption{Stellar mass functions for bol and bolmvir models (left), and
  tree and bolp models (right) for
  all galaxies at higher redshifts $\sim 0.6,  0.9$ and $2$.  Lines and
  points as in Fig.~\ref{fig:smfz0}.
The agreement of the bol model with observations at redshift $\sim 0$ diminishes
at higher redshifts, 
rising above the observations at low stellar
mass; in contrast the tree model has excellent overlap at $z\sim
0.9$. 
As redshift increases, the models sharing cosmology but differing in mass definitions and
mass gain models become closer.  At higher stellar mass
the observations tend to lie between the 1 and 2 progenitor models
until $z\sim 2$, where all the models tend to overlap, and lie at or slightly
above the observations (Table \ref{tab:smfobs}).  
}
\label{fig:smfall}
\end{figure*}

At higher redshift, Fig.~\ref{fig:smfall}, the agreement of the bol stellar mass function 
with observations decreases, while the
agreement with observations of the tree stellar mass function varies.
The tree model, with instantaneous $M_{\rm fof}$, tends to be higher
than bolp at low redshift, but then falls below it at $z=0.9$ and
closer to
the observations.   The
bolpmvir model, based upon instantaneous $M_{\rm vir}$ and not shown, has stellar mass
functions which are closer than bolp to those of the
tree model but with a slightly different shape than the tree model.  
It is hard to
interpret the changes between the tree and bolpmvir model 
as not only does the average relation between these two mass
definitions evolve with redshift, but the details of physical mass
gain are
likely to differ between these definitions as well.

At $z\sim 2$ the stellar mass functions of the simulated galaxies lie
at or slightly
above the observations for all models shown,
especially near the break in the stellar mass function.  The models overlap
closely
for all models sharing the same cosmology, including those not shown, aside
for a small variation at high stellar mass due to 1 or 2 progenitors
contributing final stellar mass.

\begin{figure*}
\begin{center}
\resizebox{2.9in}{!}{\includegraphics[clip=true]{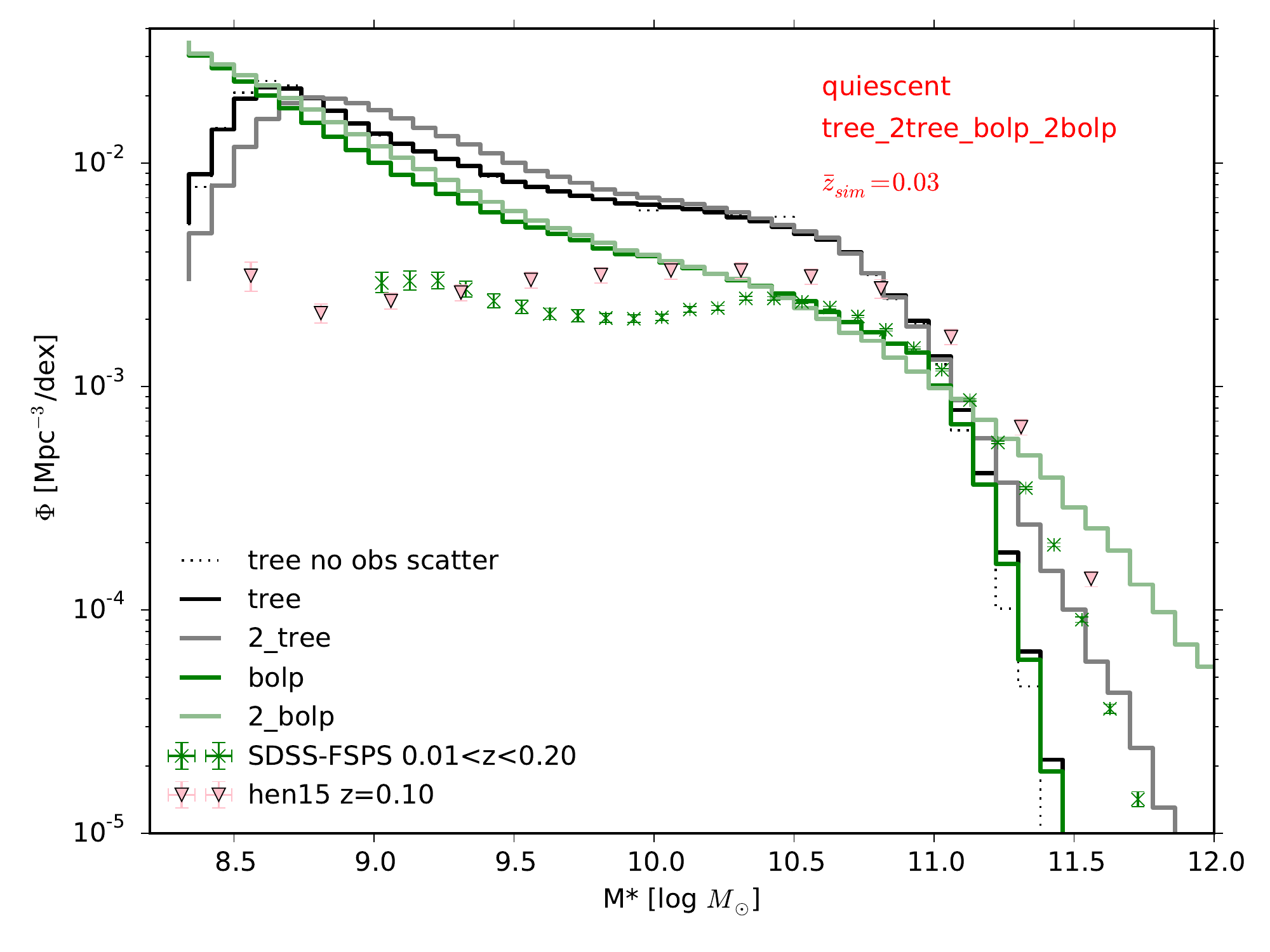}}
\resizebox{2.9in}{!}{\includegraphics[clip=true]{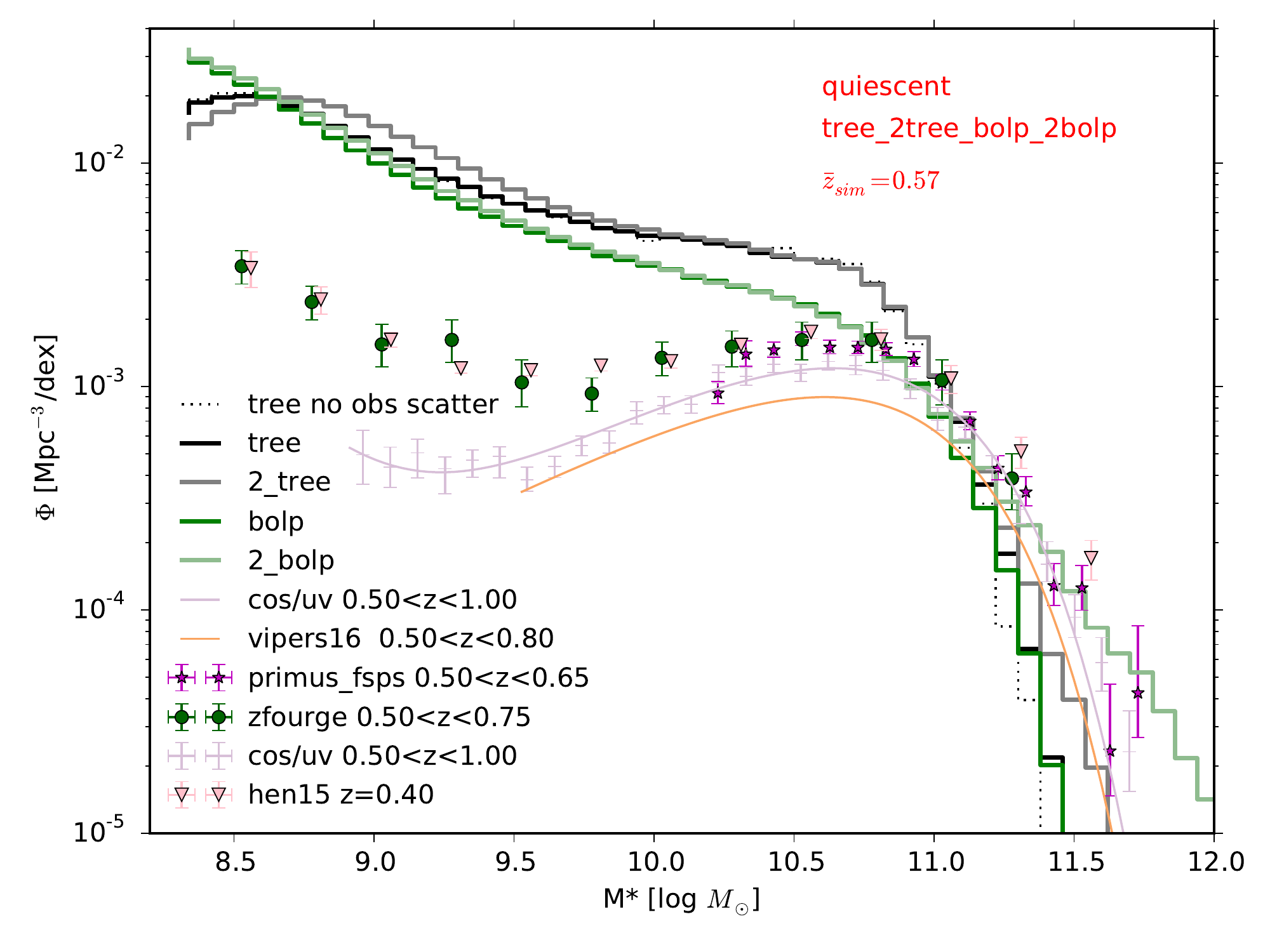}}
\resizebox{2.9in}{!}{\includegraphics[clip=true]{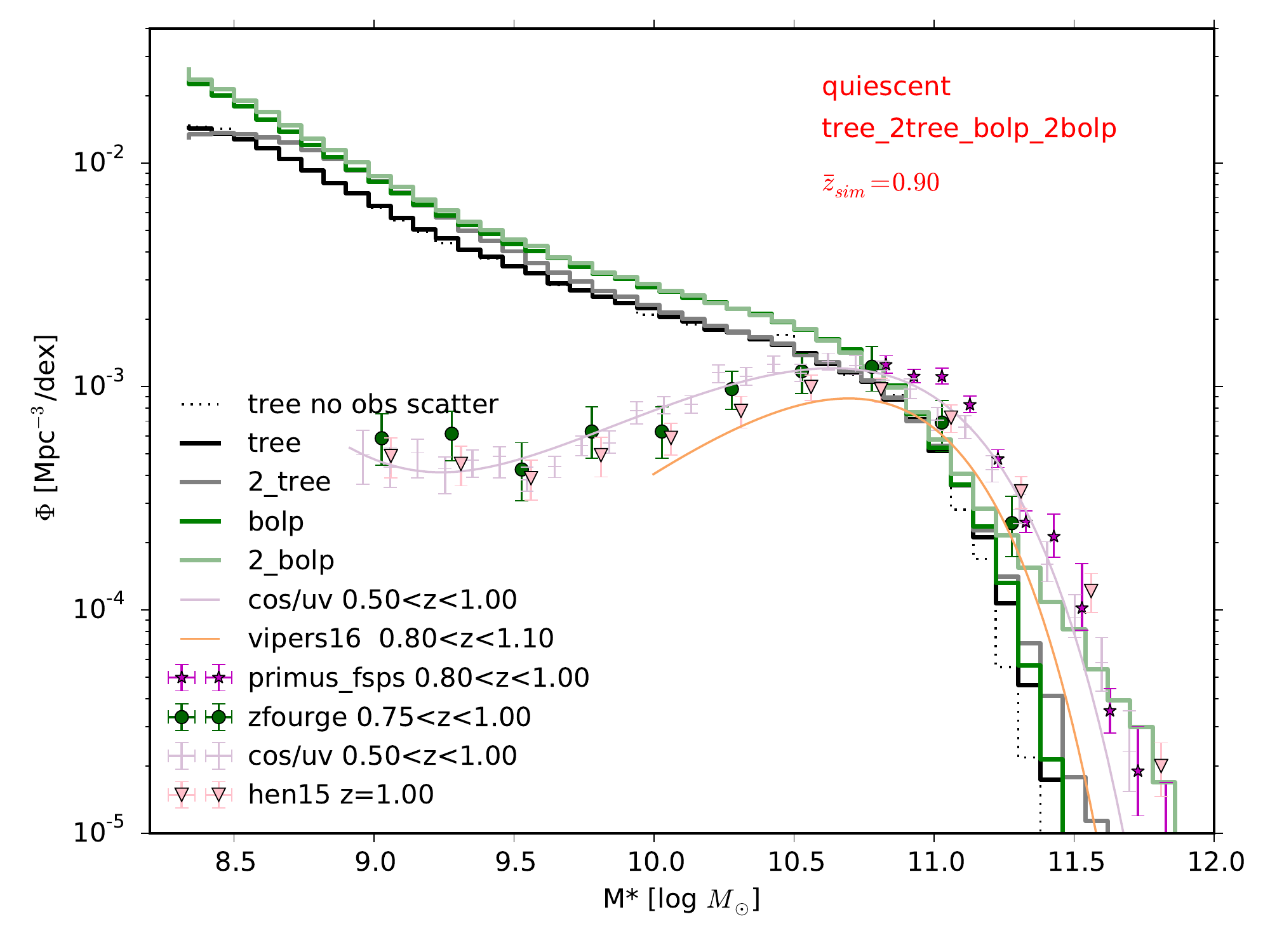}}
\resizebox{2.9in}{!}{\includegraphics[clip=true]{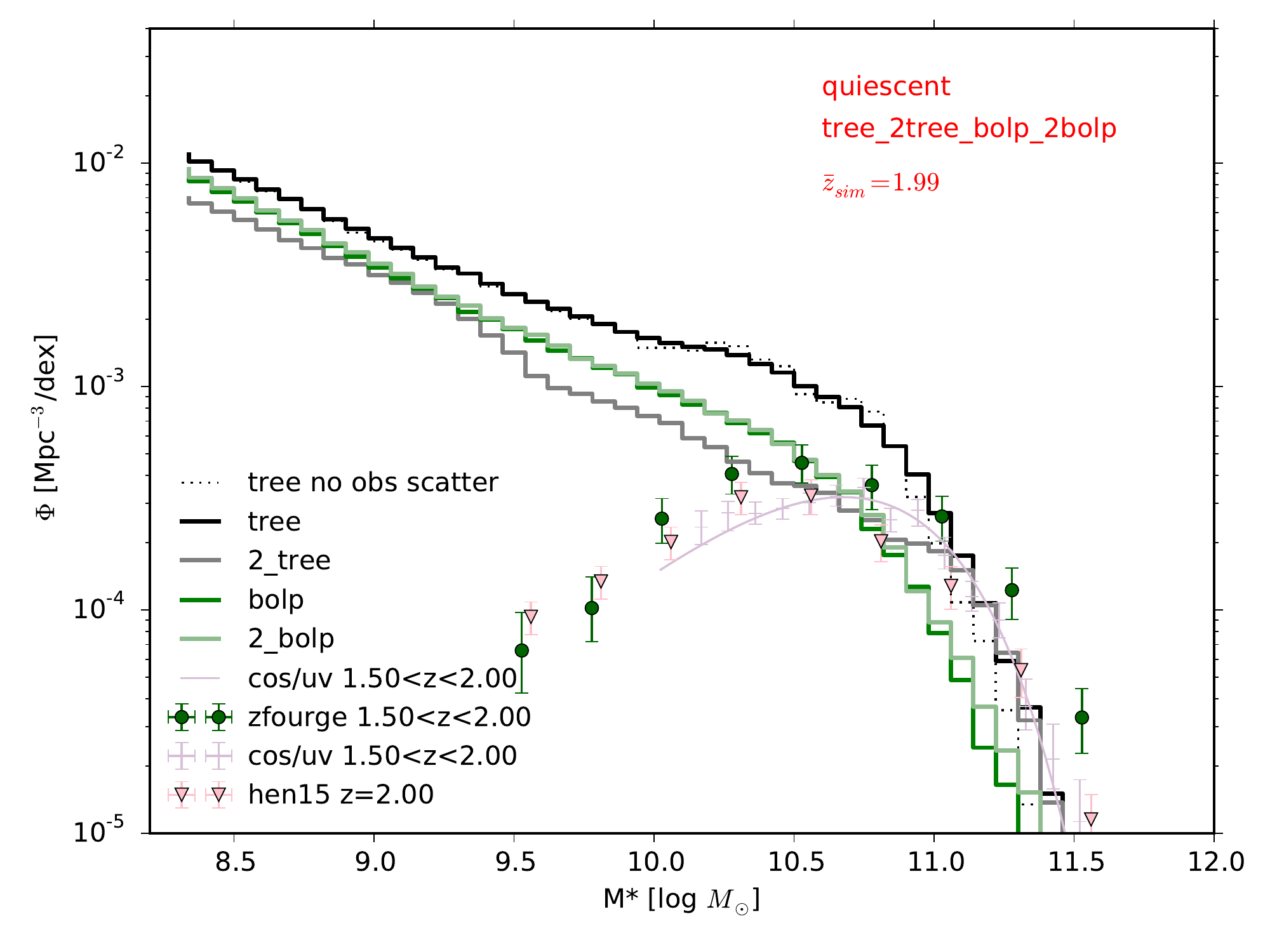}}

\resizebox{2.9in}{!}{\includegraphics[clip=true]{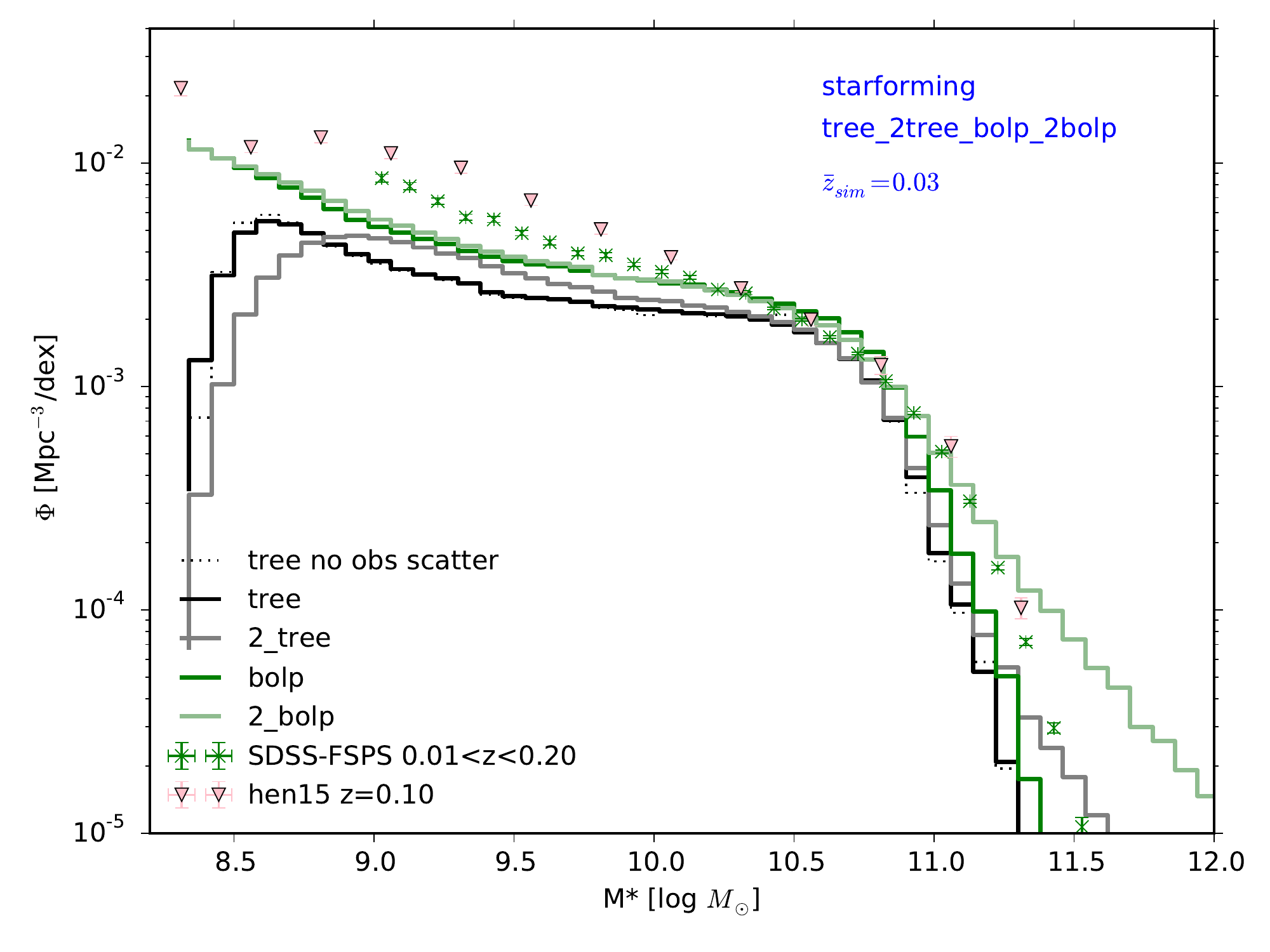}}
\resizebox{2.9in}{!}{\includegraphics[clip=true]{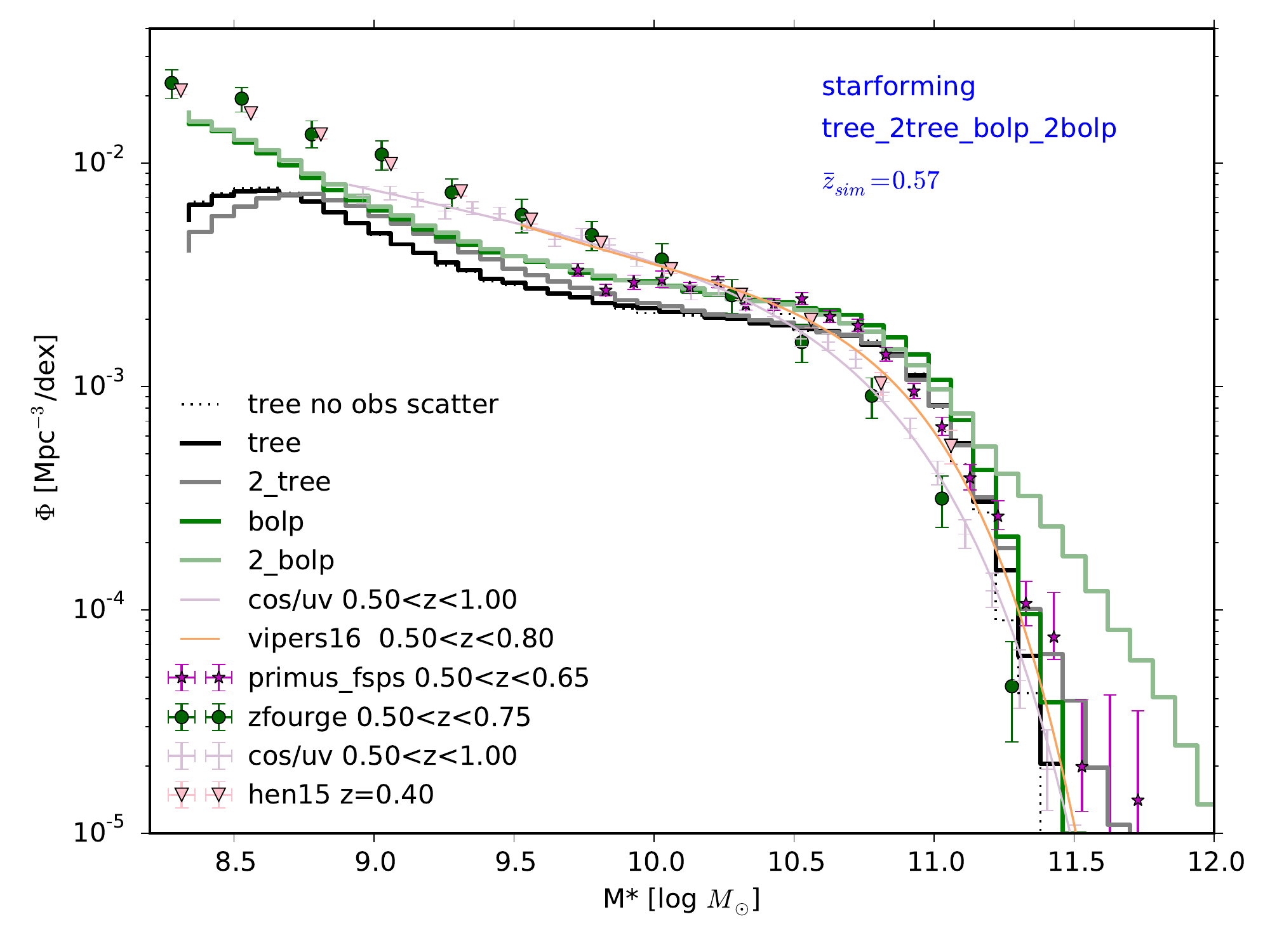}}
\resizebox{2.9in}{!}{\includegraphics[clip=true]{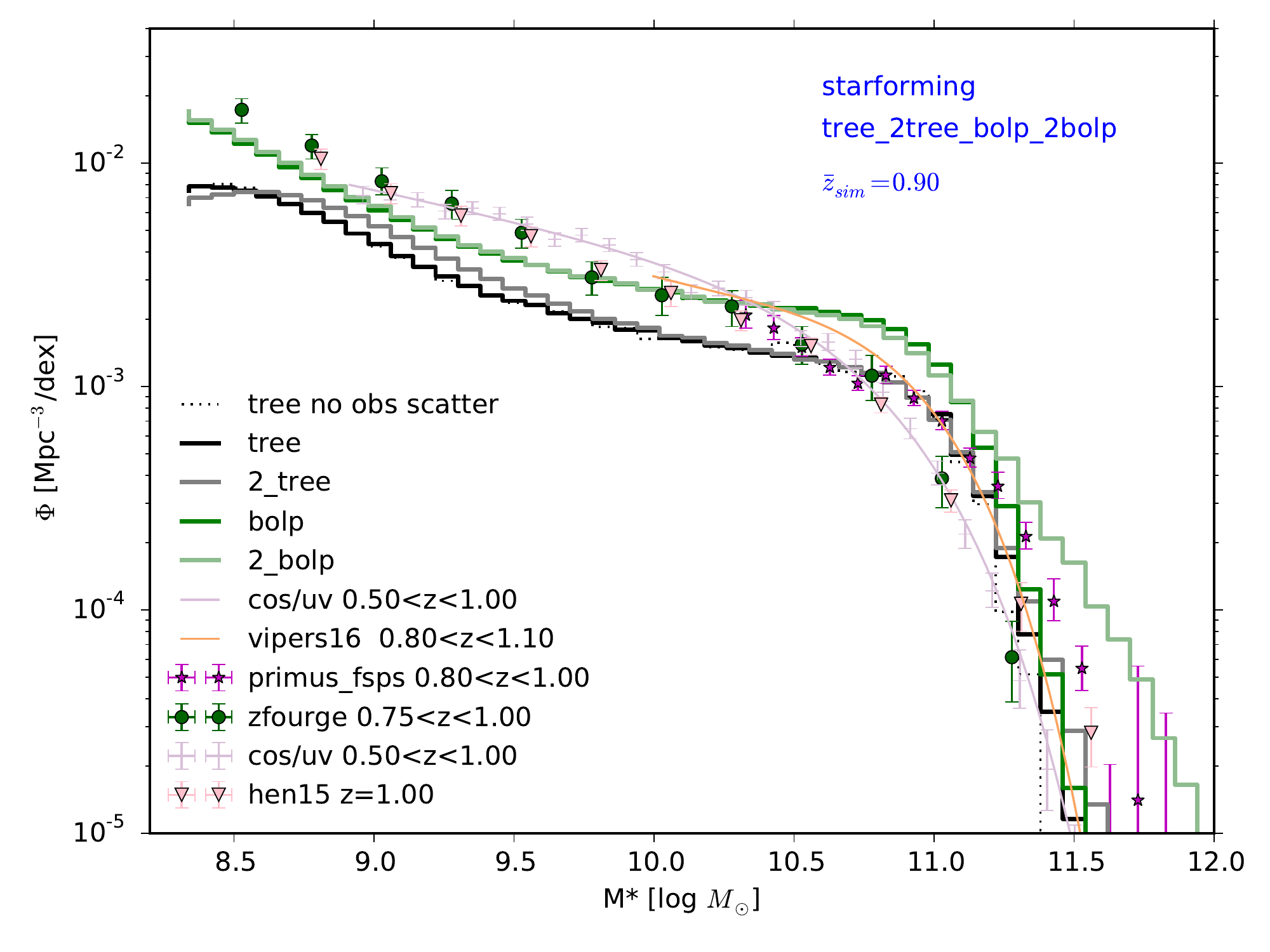}}
\resizebox{2.9in}{!}{\includegraphics[clip=true]{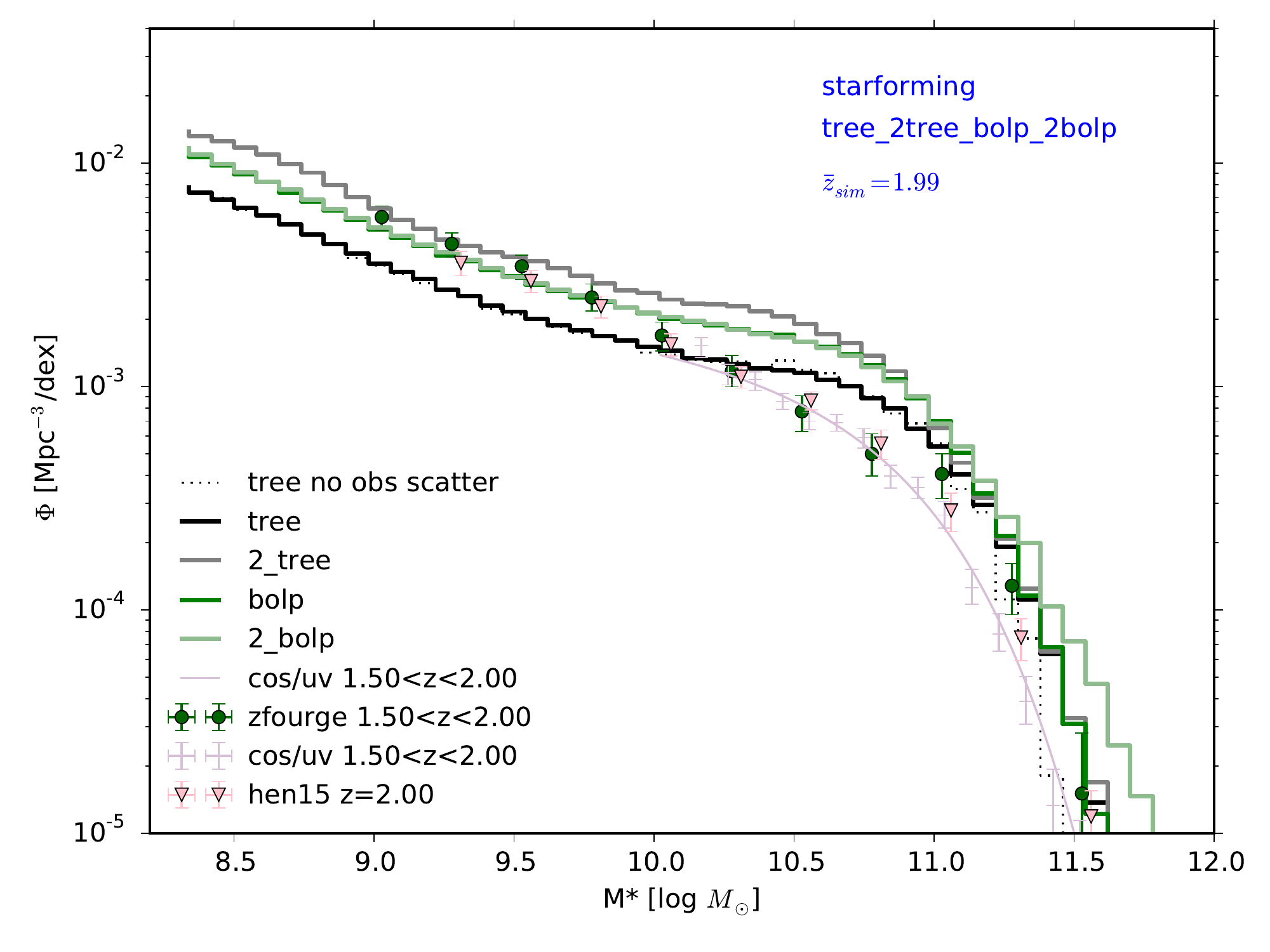}}
\end{center}
\caption{Redshift $\sim 0, 0.6, 0.9, 2$ quiescent stellar mass
  functions, top, and star forming stellar mass functions, bottom,  for
  tree and bolp. The division between quiescent and star forming is shown in
  Fig.~\ref{fig:mstarsfr}.  The observational data are described in Table \ref{tab:smfobs},
  note different observations have different definitions of
  starforming and quiescent.   The excess of quiescent galaxies for tree results in it having a 
  deficit of star forming galaxies, all models have too many faint
  quiescent galaxies.
  Other features as in Fig.~\ref{fig:smfz0}.}
\label{fig:smfcolor}
\end{figure*}

An estimate can be made
of the quiescent and star forming stellar mass functions by using the
division between the populations in terms of star formation rate and
stellar mass seen in \S\ref{sec:mstarsfr}.  
Again, the caveats about different definitions of quiescent and star
forming in the observations should be kept in mind (Table \ref{tab:smfobs}).
For comparison of star forming and quiescent stellar mass
functions,
the tree and bolp models will
be shown, as they are based upon simulations closer to the
current best fit parameters.  Their quiescent and star forming
stellar mass functions
are shown
at 4 different redshifts in Fig.~\ref{fig:smfcolor}. 
In comparison to the observations,
the simulated models give too
many quiescent galaxies at low stellar mass. 
For the tree model, some of the quiescent galaxy excess is built in.  The
tree satellite subhalos only increase their subhalo mass from their
infall mass through merging, and so tend to only have the random
added by hand component of 
star formation (in Eq.~\ref{eq:dmstardmh}).  As a result, the tree 
model has only
small fraction of star forming satellites
 (about 1 percent).  This almost automatic satellite quenching does
not occur the Bolshoi and Bolshoi-P based models,
which have much larger
fractions of star forming satellites.  
However, all of these variants show an excess of quiescent galaxies
at low stellar mass (also for the bol, etc., simulations, not shown).
This may be made even
worse if the center of the quiescent branch (a feature chosen by hand) 
is moved to a higher specific star formation rate to agree with
observations, as suggested by the stellar mass-star formation rate
diagrams in \S\ref{sec:mstarsfr}.

To summarize, the model upon which the star formation efficiencies were tuned
has a good match to the low stellar mass component of the 
stellar mass function at $z \sim 0$.  For higher stellar mass,
 the 1 progenitor and 2 progenitor models tend to separate and
the observational data usually falls between the two, 
suggesting the use of some combination of the two
models which generalizes the
prescription of \citet{Bec15}) to arbitrary time steps, mass
resolution, and perhaps including redshift dependence.
There seems to be an excess of low stellar mass quiescent galaxies for
all models.
At the highest redshift considered, $z\sim 2$, the stellar mass
functions of all the models tended to become degenerate, aside from 
differences due to cosmology, and a
small change if 1 or 2 progenitors contributed stellar mass to a
galaxy.  The models lie mostly above the
observed stellar mass function at $z\sim 2$.
\begin{figure*}
\begin{center}
\resizebox{3.2in}{!}{\includegraphics[clip=true]{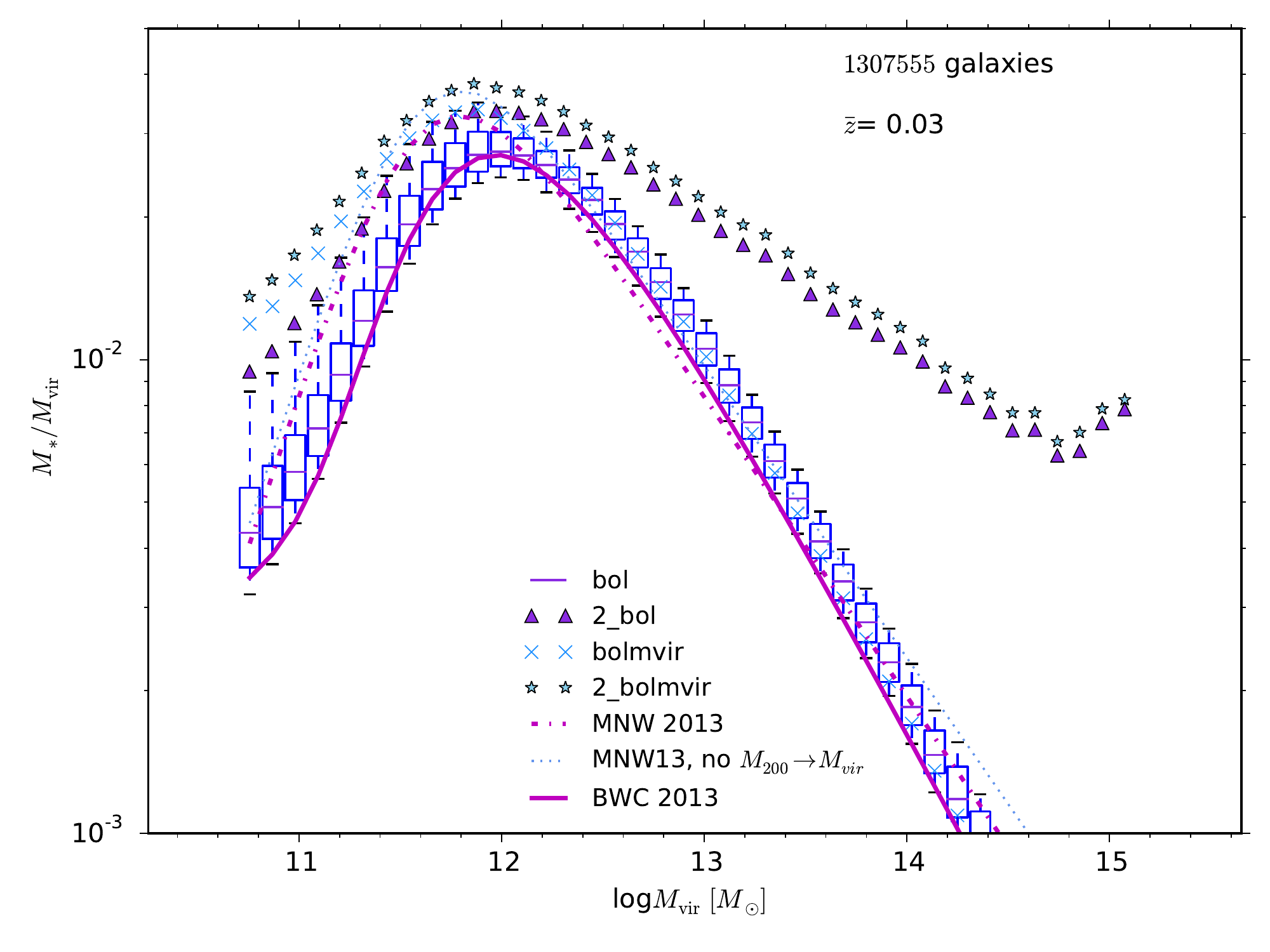}}
\resizebox{3.2in}{!}{\includegraphics[clip=true]{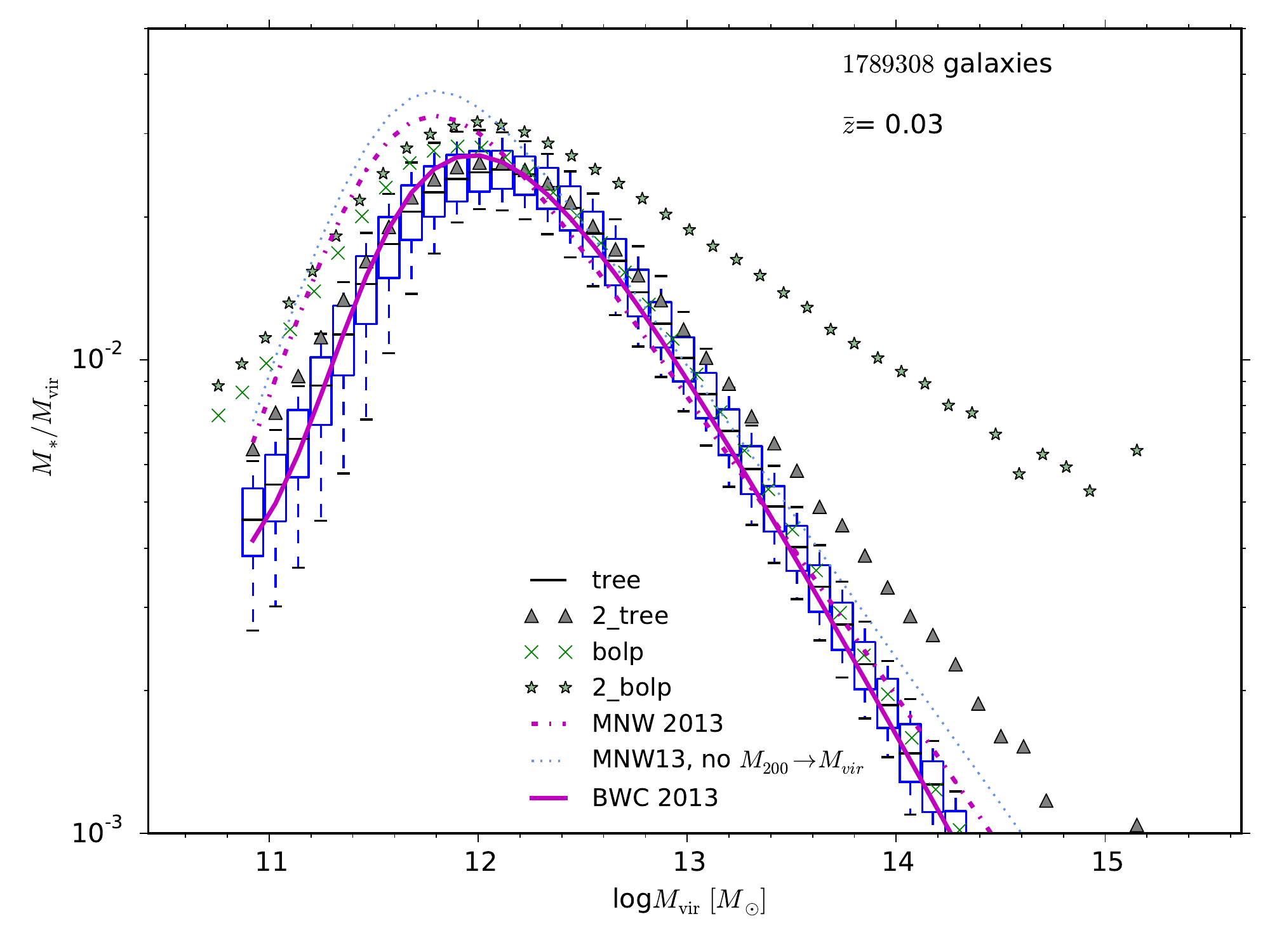}}
\resizebox{3.2in}{!}{\includegraphics[clip=true]{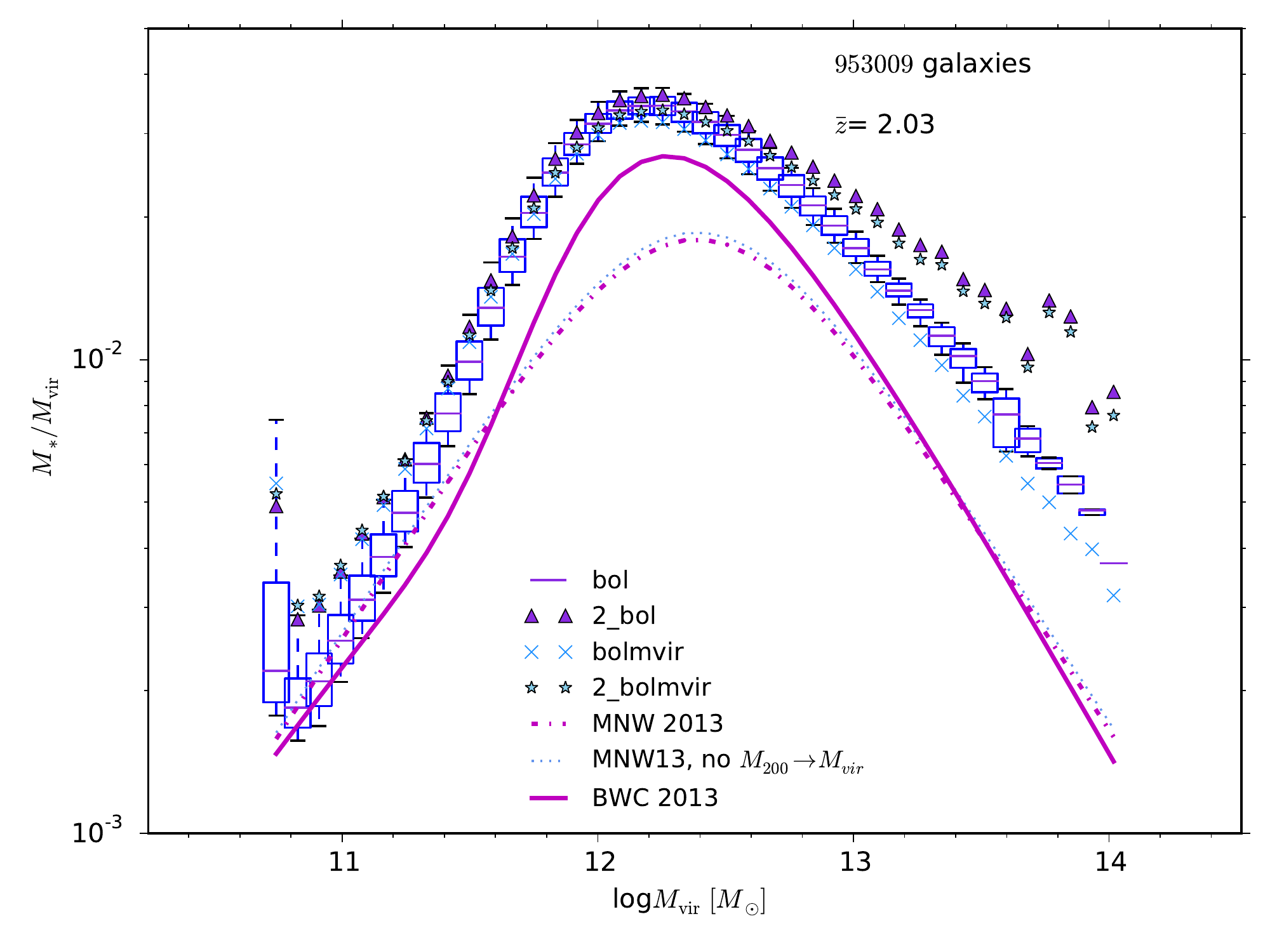}}
\resizebox{3.2in}{!}{\includegraphics[clip=true]{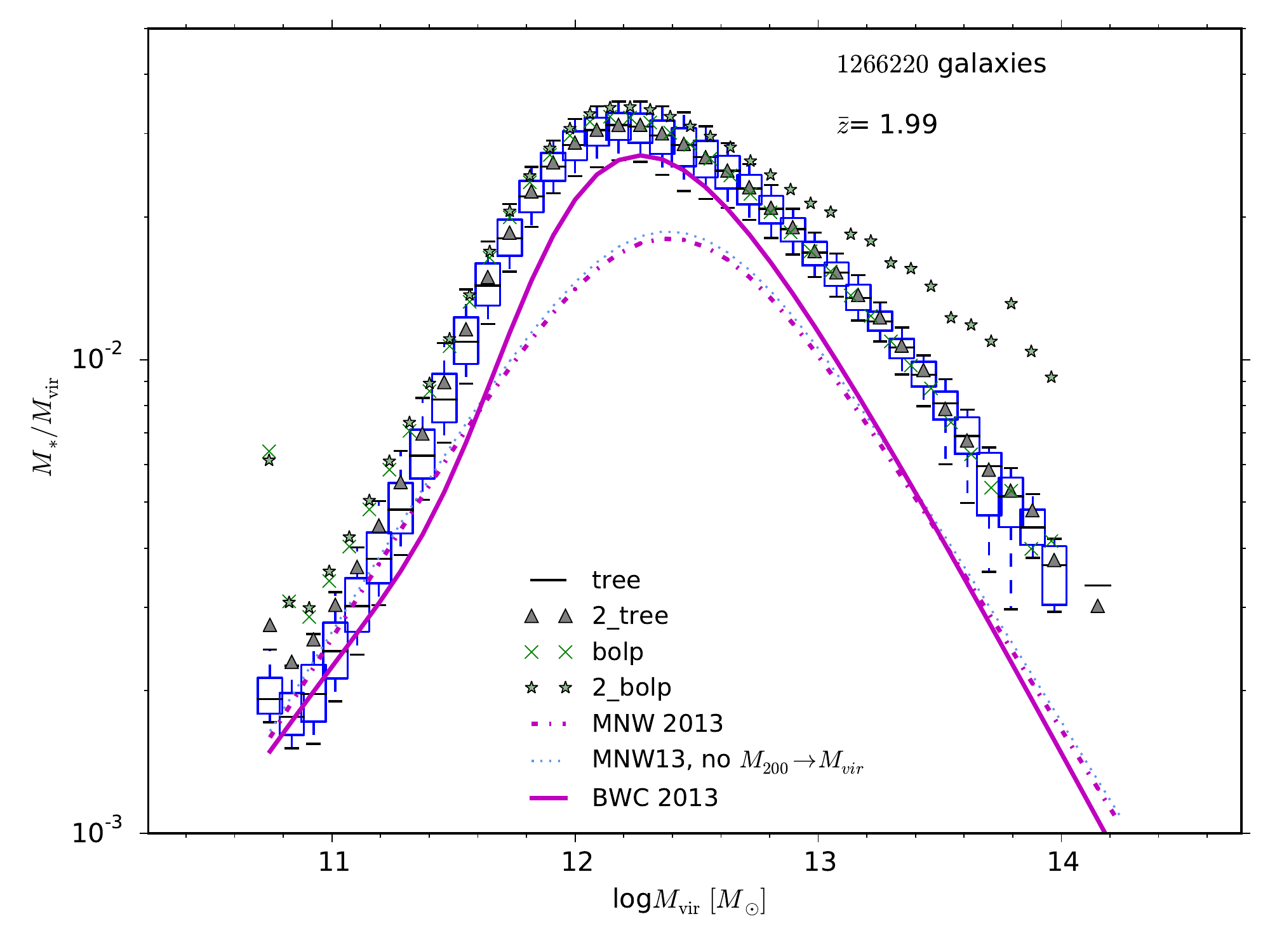}}
\end{center}
\caption{$M^*(M_h)/M_h$ as a function of halo mass $M_h$ for low
  redshift (top) and high redshift ($z\sim 2$, bottom), for the bol,
  bolmvir models at left and the tree, bolp models at right.
The number of galaxies listed are for the 2bolmvir and 2bolp models, respectively.
The two lines for \citet{MosNaaWhi13}
  correspond to the formula directly in their paper (dotted) and
  converted
to $M_{\rm vir}$ (dot-dashed) and the solid line is from \citet{BWCz8}.  The scatter
in the measured ratios (not shown) is in the range 30\%-50\% for most
values of $M_{\rm vir}$.  The boxes are for the 1 progenitor bol or tree models,
showing 
upper and lower quartiles around the
median value, shown by a line, and the ``whiskers'' are at 10 and 90
percentiles.  For clarity,
only the median value for the 2 progenitor bol and tree (triangles) models, the 1
progenitor (X's) and 2 progenitor (stars) bol and bolmvir
models are shown.
Galaxies with $M^*<10^8 M_\odot$ were not included; the low $M^*$
cutoff can introduce spikes at low $M_{\rm vir}$.
}
\label{fig:msmh} 
\end{figure*}

\subsection{Stellar mass to halo mass relation}
\label{sec:smhm}
A third comparison quantity between the models and observations is
how a galaxy of a given stellar mass inhabits a dark
matter halo, $M^*(M_h)$.  For the simulation based models, this
relationship is immediately available.

In Fig.~\ref{fig:msmh}, the 
curves show two measurements of $M^*(M_h)$
from observations,
by \citet{MosNaaWhi13} and \citet{BWCz8}.  (The functional
forms of the curves are in Appendix \S
\ref{sec:appmsmh}.)  
As the simulated halo masses
are $M_{\rm vir}$ halo masses, and \citet{MosNaaWhi13} uses $M_{200c}$
masses, the \citet{MosNaaWhi13} curve is shown both with and without
an applied average conversion \citep{Whi01} to the halo masses, 
to give an idea of the size of the conversion effect.  
These $M^*(M_h)$ fits differ in the assumed
cosmology and also in the observations used to derive them.   Both
use \citet{PerGon08}, while \citet{MosNaaWhi13} also uses \citet{San12},
and \citet{BWCz8} also use \citet{Mou13} in the redshift range $0-1$.
The observational stellar mass functions for both fits have
the \citet{BC03} stellar population synthesis models with a Chabrier
IMF, and so are expected to be comparable to the \citet{Bec15} models here,
which has both of these assumptions incorporated into the star
formation efficiencies.  However, the dust model of
\citet{MosNaaWhi13} might differ.
Although the two theoretical curves look different, errors for
\citet{BWCz8}, given in, e.g., the downloads
sfh\_z0\_z8.tar.gz available at www.peterbehroozi.com,
 are often large enough to encompass the average $M^*(M_h)$ for
\citet{MosNaaWhi13}.

The \citet{BWCz8} and \citet{MosNaaWhi13} curves are compared to 
the bol, bolmvir, tree and bolp simulated 1 and 2 progenitor models in
Fig.~\ref{fig:msmh}, at redshift 0 and 2.
For simplicity, only the 1 progenitor bol or tree model is
shown in detail.
In this case, the line in the 
center of every box is the median, and the boxes go from lower to upper
quartile of the data. 
The solid black lines above and below are 10th to 90 percentile.  For
the
other 3 models, only the median values are shown:
triangles are for the 2 progenitor tree or bolp models, 'X's'
for the 1 progenitor bolp and bolmvir models and stars for
the corresponding 2 progenitor
 models.  
 The example shown in detail shows that the median tends to lie near the
 average of the distribution except towards high halo mass.  All galaxies are included here, not just central galaxies.
As there is large scatter in the measured $M^*(M_h)$, roughly a factor
of two for BWC13 for example (not shown for clarity),
the lack of exact agreement with the medians of the distributions
in the simulations is not a concern, although all of the model distributions
look high at redshift $\sim 2$.  However, the 2 progenitor
models at high stellar mass seem to put galaxies
in halos with not enough mass (except for the 2tree model at high redshift).
Again, this suggests that a combination of 1 and 2 progenitor models
may agree with observations. 

To summarize the comparison between the models and observations,
the simulated galaxies have stellar mass-star formation rate relations
which are clearly bimodal, but which seem to have star formation rates
which are low, on the whole, compared to \citet{Mou13}, and the
difference increases with redshift.  This may be in part due to
difference in stellar mass definitions and the by-hand choice of the
position and scatter of the quiescent branch.
The models based upon instantaneous halo mass (tree, bolmvir, and
bolpmvir) tend to have more features in their stellar mass-star
formation rate diagrams, relative to 
those built upon the smoothed and constrained $SAM\_M_{\rm vir}$.
In the specific star formation rate, the models are similar, although 
the smoothed $SAM\_M_{\rm vir}$ models sometimes do not have as
clear a break in specific star formation rate between the star forming
and quiescent galaxies.  The tree models have the largest number of
quiescent
galaxies, in part because almost all of the satellites are quiescent.

The observed stellar mass function at low redshift agrees best with
the simulation which was also used to find the
star
formation efficiencies as a function of mass (upon which the model
here is based).  At high redshift $z\sim 2$, all of the simulations have stellar
mass functions which overlap, for variants which have the same
cosmology (aside from some variation at high stellar mass between
to 1 or 2 progenitor versions), and
all have some overlap with observations, although the simulations are 
high near the bend
in the stellar mass function.
There seems to be an excess of faint quiescent galaxies
at all redshifts in all models.
The stellar mass to
halo mass relations are mostly reasonable, although they tend to be
higher than the central values found observationally (albeit based on
a now disfavored cosmology), and the 2 progenitor
models are far from observational constraints,
especially at high halo mass and low redshift.
It seems some combination of 1
and 2 progenitor models, with the criteria flexible enough for different
time steps and halo mass resolution, may be able to reach the
observational values falling in between the two variants for the
stellar mass to halo mass relation.  Another possibility is to lower
the stellar mass per halo via more stellar loss mass into the ICL,
even from just one progenitor.

\section{Ensemble of galaxy formation histories}
\label{sec:histories}
The models constructed here are very simple when compared to many others
found in the literature and yet seem to have reasonable observational
properties.  It is thus interesting to see how they compare to
semi-analytic models, which use and evolve many more physical halo and
galaxy properties, and give many more predictions.
One way to compare different galaxy formation models 
is to look at the ensemble of histories of galaxies they produce.
Properties of the histories of galaxy formation, 
for a large number of galaxies in a
model, is another way to characterize how galaxies form in that model.
(Individual galaxies can only be followed in detail only in
simulations.\footnote{Methods to deduce histories with increasing
  detail from observational spectra are also being
developed, see, e.g., \citet{Pac13,Pac16}.})

In this section the full ensemble of galaxy 
histories for several variants of the \citet{Bec15} model are compared in terms of their average histories,
the variance around the average history, the leading fluctuations
around
the average history and the fraction of variance in several of the
(leading)
fluctuations around the average history.  These quantities are readily
available using principal component analysis (PCA).
Using PCA, 
the ensembles of histories of stellar masses, in both a semi-analytic
model and a hydrodynamical model,
were found to be well approximated as combinations of only a few basis
fluctuations around the average history in \citet{CohVdV15}.  
In this section, histories of the simple models constructed above 
are compared to histories of
the \citet{Guo13,Hen15}
semi-analytic models built upon the Millennium \citep{Spr05,Lem06} 
simulation, which have many more parameters and detailed predictions,
and to a straw man model based solely upon subhalo mass.

\begin{figure}
\begin{center}
\resizebox{3.5in}{!}{\includegraphics[clip=true]{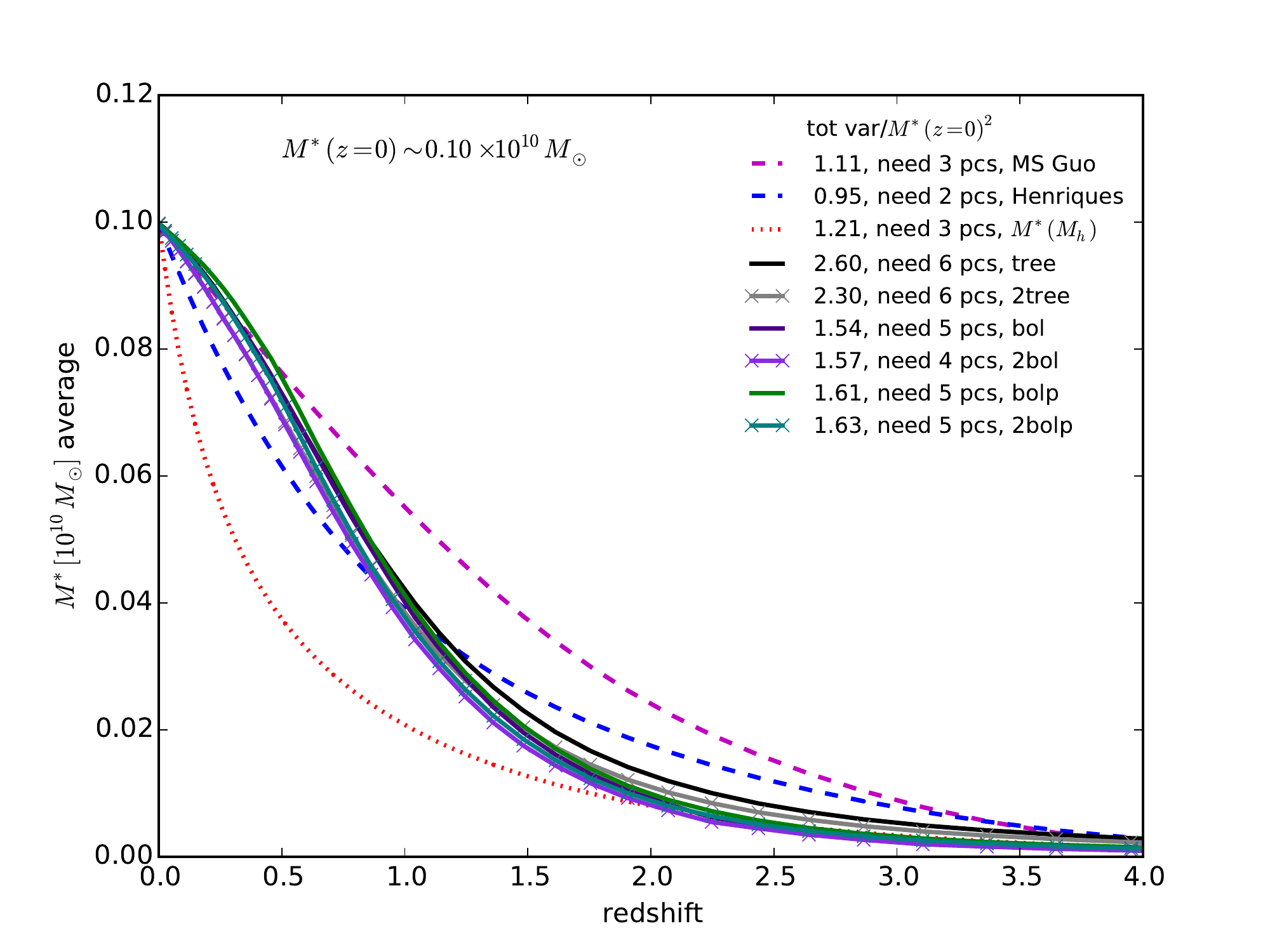}}
\resizebox{3.5in}{!}{\includegraphics[clip=true]{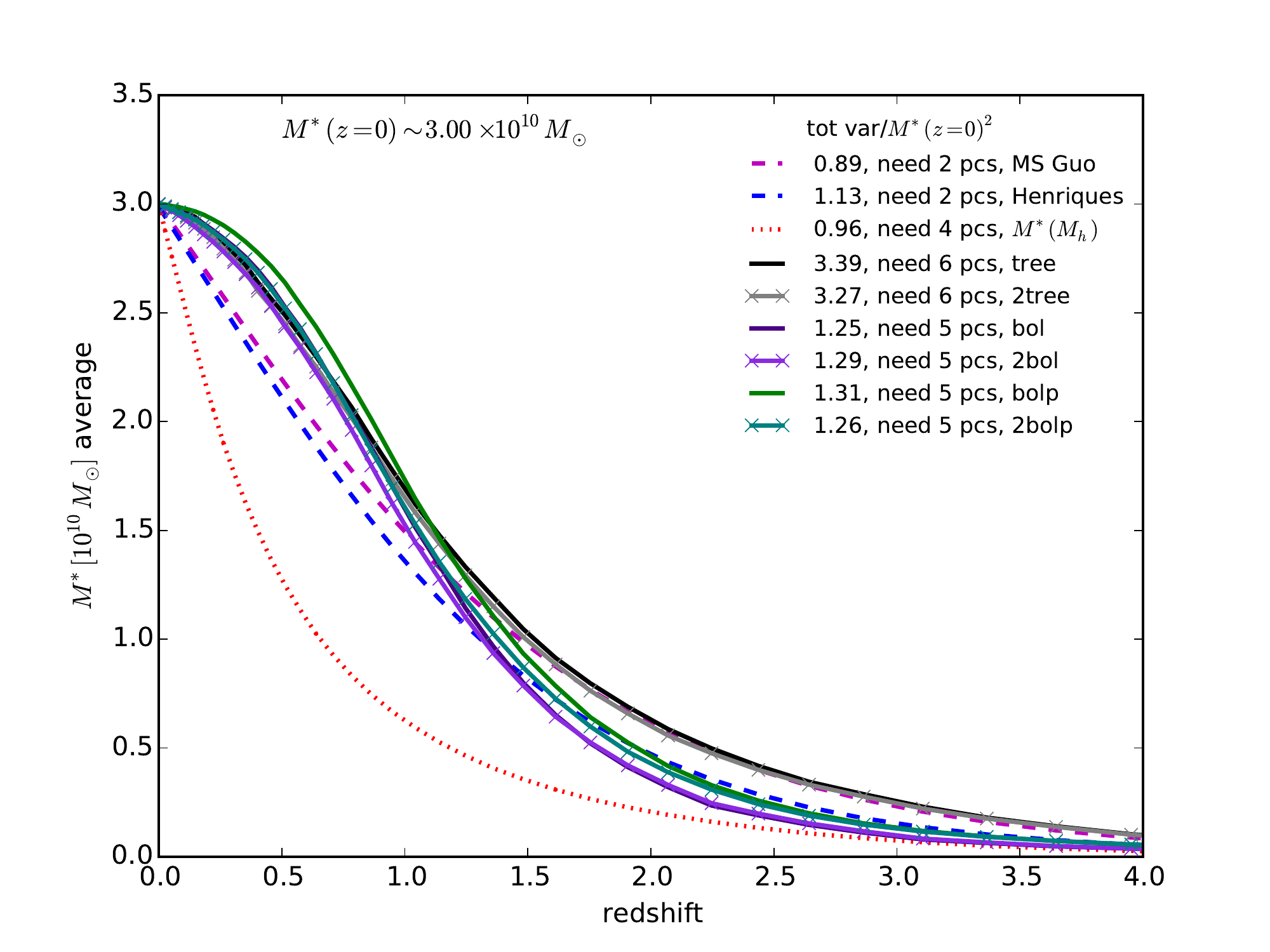}}
\resizebox{3.5in}{!}{\includegraphics[clip=true]{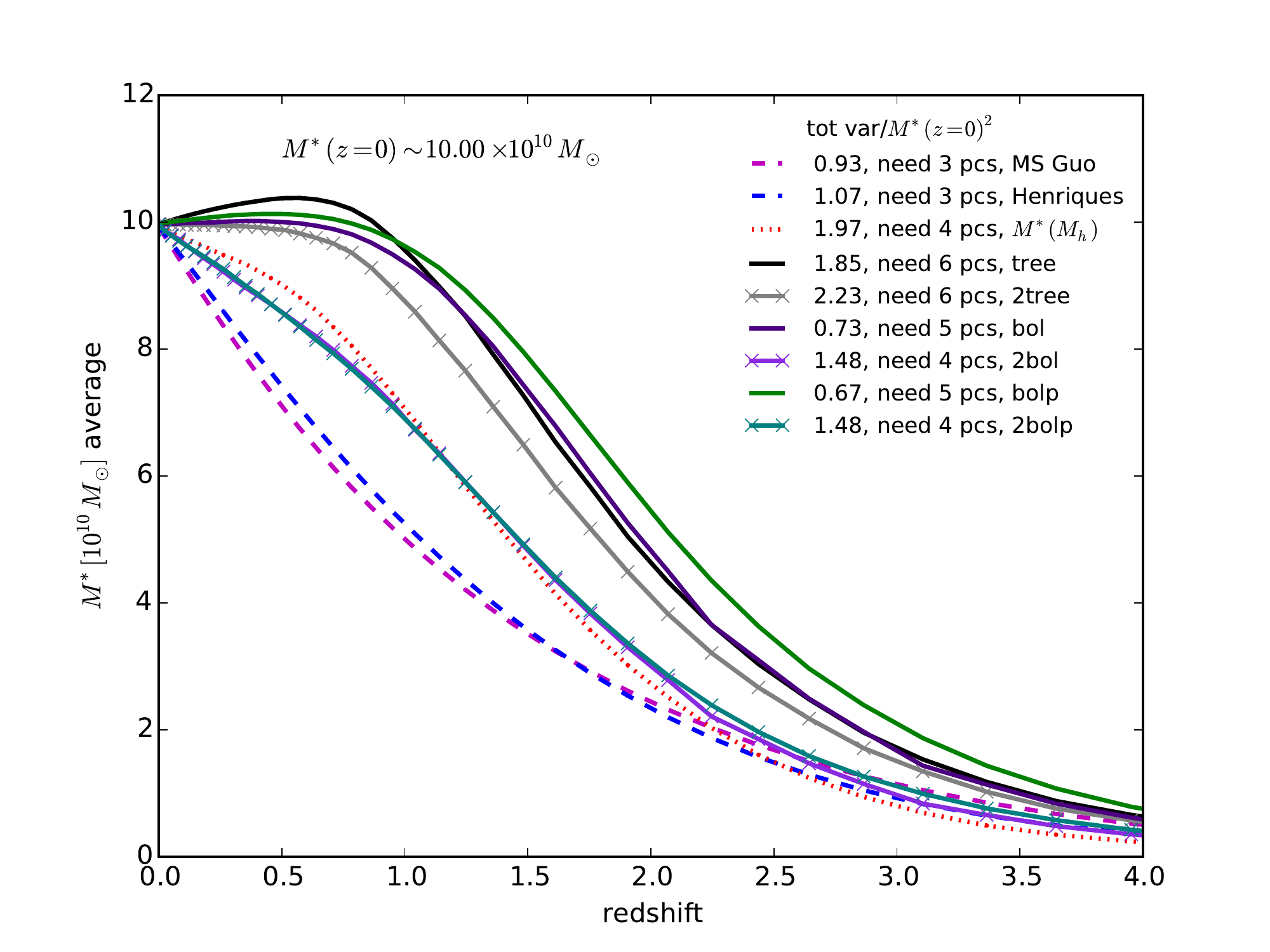}}
\end{center}
\caption{Comparison of average stellar mass histories for 9 different models built on dark matter merger
  tree histories, for galaxies which have $M^*(z=0) = 10^{9}, 3\times
  10^{10}$ and $10^{11}
  M_\odot$, top to bottom.  The
models are described in the text. 
The basis fluctuation contributing the largest variance for
each of the models tends to peak at earlier times for average
histories with earlier stellar mass gain, and to be similar for models with
average histories which are similar.
The legend shows the normalized total
variance/$M^*(z=0)^2$, and number of principal components (out of 38)
needed to get 90
percent of the variance.}
\label{fig:comphist3} 
\end{figure}

To implement PCA,
the stellar mass history of every galaxy in a simulation is taken to be
a vector, with the stellar mass at each time corresponding to a different
component of the vector.
One can then study and classify the properties of the distribution of
these vectors for a given set of histories, i.e. instance of a 
galaxy formation model.  Principal component
analysis (PCA), mixture models and k-means clustering were all
considered in \citet{CohVdV15}.  Here PCA is used, as it gives
several quantities to compare besides the separation of 
histories into classes.  
PCA is applied by finding the covariance matrix of the different
components of the vector of stellar mass histories and diagonalizing it.
In \citet{CohVdV15}, the model 
histories studied from \citet{Guo13} and \citet{OWLS} had
fluctuations around the average stellar mass history
which could be described in large part by only 3 basis fluctuations
times
some coefficients.  (In more detail, galaxies compared were taken to have
approximately
the same final stellar mass, and 
$\sim$ 90\% or more of the variance around the average was captured by
3 basis fluctuations out of 42).
The PCA approach thus compactly describes 
the full ensemble of stellar mass histories in a model, for any given final stellar
mass.\footnote{
Histories of star formation rates require
more basis histories, often close to half of the total number of
histories, to span the same fraction of
their variance.  Histories of dark matter halo masses were studied in
\citet{WonTay11}.}

One of the models from \citet{CohVdV15} is considered here, for
comparison,
the
\citet{Guo13} model built upon the Millennium \citep{Spr05,Lem06}
simulation.
A second updated model, \citet{Hen15}, also built on the Millennium
simulation, is used to illustrate the variations found for this method 
between reasonable semi-analytic models. 
Along with these, 
the 1 and 2 progenitor varieties of the
tree, bolp and bol models are analyzed using PCA.
(As a reminder, galaxies in both 1 and 2 progenitor 
variants of the \citet{Bec15} model 
gain stellar mass from star formation based on their accreted 
halo mass, however, for the 1 progenitor variant, only the stellar mass of the
most massive progenitor is inherited by a galaxy, while in the 2
progenitor 
variant the
stellar mass of the two most massive progentors is inherited.)  
In addition, 
a straw man model is considered, where stellar mass is completely
determined by
$M^*(M_h,z)$ of \citet{MosNaaWhi13}. 
That is, every (sub)halo in the TreePM simulated trees is required to lie exactly on the average $M^*(M_h)$
relation at every time step (although the relation came from abundance
matching, this is not abundance matching per se, as scatter is not included).  
For this straw man model, satellites are assigned $M^*(M_h,z)$ at infall redshift $z$, 
and have fixed stellar mass afterwards. 
The \citet{MosNaaWhi13} $M^*(M_h)$ relation is given by Eq.~\ref{eq:mosmsmh} in the appendix, 
and shown in Fig.~\ref{fig:msmh} for a few
redshifts.  As this relation requires $M_{\rm 200c}$ masses, the FoF
masses
are converted via \citet{Whi01}.  
(This $M^*(M_h)$ model is related to one presented in
\citet{MosNaaWhi13}, which there included scatter.)

As in \citet{CohVdV15}, the histories are studied for small ranges of
final stellar masses.  In that study, a qualitative change in the 
history properties occurred around final stellar mass 
$M^* \sim 3\times 10^{10} M_\odot$.  Thus,
here $\sim 10^{9}
M_\odot$, $\sim 3\times 10^{10} M_\odot$, and $\sim 10^{11}
M_\odot$ are chosen as final stellar masses.\footnote{More precisely,
  the range was roughly 9.5-10.5 $\times 10^{9}
  M_\odot$, 2.95-3.05 $\times 10^{10} M_\odot$ and 9.5-10.5 $\times 10^{10}M_\odot$
  in final stellar masses.  Altering ranges by a small amount 
  did not seem to change the results significantly.}
 
The different underlying dark matter simulations have different numbers of time steps.
The
  histories are followed from redshift $\sim 4.5$ to $\sim 0$, which
  takes 42 steps in the Millennium simulation, 38 in the Millennium
  Planck simulation, 91 in the TreePM
  simulation, 165 in the Bolshoi simulation and 142 in the Bolshoi-P 
simulation.  Doing PCA on the
  different histories as given, with their different time steps and
  numbers of time steps relative to Millennium, leads to much larger scatter in
  the models explored in this paper, which all have many more time
  steps than the Millennium simulation.  
These model histories are thus interpolated to
  the 38 values where the Millennium Planck based model has outputs, to
  make the scatter in the histories more directly comparable.  (This
  increase of scatter with number of steps was not as evident when
  comparing the hydrodynamic simulations, with 20 time steps, 
with the \citet{Guo13}
  stellar mass histories, with 42 time steps, 
in \citet{CohVdV15}. ) Other properties, such as the shape of the
leading fluctuations, or the number of fluctuations, did not seem to
change as much as the number of steps was changed.
These interpolated
  histories with 38 time steps are the ones compared in the figures.  In addition,
  histories where any step had more than a 30\% stellar mass drop were taken
  out, which varied from up to 1 percent in the 2bolp model to 0 in the semi-analytic
  models.

Some results of PCA analysis of these stellar mass ranges for 
these nine models are shown in Fig.~\ref{fig:comphist3}. 
  Each figure (corresponding to a different final stellar
mass) shows the average histories for galaxies in each of the 9 models,
from redshift 0 to
4.3.  The number of $PC_n$ (basis fluctuations around the average) needed to get 90 percent
of the variance is shown in the legend, more $PC_n$ (or in a few cases the
same) are needed for the simple models explored in this note, 
relative to the semi-analytic models.
This number of components should be compared to 
the total number of $PC_n$ fluctuation components, 38, to estimate how
well a few fluctuations capture the variance of the ensemble of galaxy
histories. (Or 37 components, as one component
is trivial because the final stellar masses in each set of galaxy histories are all rescaled to coincide
before applying PCA.)
The larger the fraction of variance due to the first few $PC_n$, the
better the histories can be approximated by only these first few
$PC_n$ times some coefficient, added
to the average history. 
The total variance for each model is listed in
the legend at right, in units of the final stellar mass squared.  The
average history and form
of the leading fluctuation, $PC_0$, are similar using either the original simulation time steps
or the interpolated ones.   Generally $PC_0$ (not shown, the fluctuation whose
coefficients have the most variance around the average history)
has a single peak which tends to be at earlier times for average
histories which have earlier stellar mass gain.
The results for each different final stellar mass are as follows.

For the lowest final stellar mass example, $M^*(z=0) = 10^9 M_\odot$,
the average histories for the \citet{Guo13} and \citet{Hen15} models
differ noticeably from each other. This is consistent 
with the fact that \citet{Hen15} models were
in part designed to have later formation of low final stellar mass
galaxies than the \citet{Guo13} model.
The $M^*(M_h)$ model has even later
stellar mass gain.  The 6 variants of the \citet{Bec15} 
models, independent of cosmology,
number of progenitors, or mass definition used, roughly collapse onto one
trajectory, with average stellar mass gain later than both
semi-analytic models.  They do not just track halo mass, although they
overlap with the $M^* = M^*(M_h)$ model at early
times.  Around redshift $\sim 2$,
 they gain stellar mass more quickly than the $M^*(M_h)$ model on
 average.  The leading fluctuation $PC_0$ around the average
galaxy histories is very similar for the 6 models constructed here, and
differs for the other 3.

Going to the second plot, for $M^*(z=0)=3 \times 10^{10} M_\odot$, 
the two Millennium models become much
closer to each other, while the 6 models constructed here 
start to split into roughly two
groups, tree and bol, bolp. 
At early times, the average
tree history is similar to the average \citet{Hen15} history,
however, at redshift 1.5 and after, the stellar mass gain for the
tree models increases sharply to join that of the other models
constructed here,
much more quickly than the two semi-analytic models or the $M^*(M_h)$
model.
 Not only do the average histories
and leading fluctuations split between tree and bol, bolp, but the variance around the
two tree models is over twice that for the other four \citet{Bec15}
based models.
All 6 models constructed here again require
more principal components to
capture at least 90\% of the variance around the average history,
relative to the semi-analytic models.  Again, the 1 and 2 progenitor
variants have very similar average histories and scatter.

For the highest final stellar mass sample, $M^*(z=0)=10^{11} M_\odot$, analyzed in
the
third set of plots,
the $M^*(M_h)$ model now has
earlier stellar mass gain on average compared to the Millennium based
\citet{Guo13,Hen15} models, which are similar to each other.  (This
similarity is expected as much of the change between the two
semi-analytic
models was aimed at galaxies with lower final stellar mass.)  There
are 
significant differences between many of the 6 models constructed using \citet{Bec15}.
The average histories for the 2
progenitor bol and bolp models 
are very close to that of the $M^*(M_h)$ model, while the
other variants gain
stellar mass much earlier.
As these high stellar mass galaxies are often
quiescent at late times, it is possible that in the semi-analytic
models these high stellar mass galaxies are inheriting stellar mass from 
even more than two progenitors to get the later stellar mass gain.  
For this larger final stellar mass, the total variance is
higher for the 2 progenitor bol, bolp models and lower for the 1 
progenitor bol, bolp models, relative to the semi-analytic models,
while both tree variants have more variance
around the average history.  All six require more $PC_n$  to capture 90\% of the
fluctuations than the 3 required for both the \citet{Guo13} and \citet{Hen15} semi-analytic models.

In summary, the average stellar mass histories and fluctuations for the 
models constructed here 
differ from those in the two Millennium based semi-analytic
models, and the $M^*(M_h)$ straw man model,  
for all final stellar mass ranges considered.  This is
independent of dark matter simulation or mass definition used, or 
how many progenitors contribute to the stellar
mass of a descendant halo.  However, the models do all have
the same star formation efficiency, so other model changes are in
principle possible.
For low final stellar mass, the different models constructed here have almost the same
 average history, and the perturbation of each corresponding to the direction of largest
variance is also very similar.
As the final stellar mass increased, these models separated out.
When they differed, the 1 progenitor models tended to have earlier stellar mass gain than
the 2 progenitor models, for fixed final stellar mass.

It is interesting that although the models are simpler in construction
than the
semi-analytic
models, they tend to require more basis fluctuations to capture most of their
variance,
and their variance around the average history, except for the 1 progenitor
models for the largest final
stellar mass, is also larger.  
This might hint at some smoothness in
the
semi-analytic models that is not captured with the simplifications
used here.
(For the tree model, some of the variance for the tree model is due to the
random star formation.  Setting the random star formation to zero 
for $M^*(z=0) = 3 \times
10^{10} M_\odot$ reduces the variance from by about 20\%, however
the number of principal components to capture 90\% of the variance
is unchanged.\footnote{I thank M. White for suggesting this test.})

\section{Discussion and conclusions}
\label{sec:conc}
The \citet{Bec15} model assigns stellar masses and star formation
rates to subhalos in a dark matter simulation, primarily using halo mass gain to
determine stellar mass gain.  Galaxy distributions and histories were
calculated for three different dark matter simulations, with differing
cosmologies, mass definitions, mass history constructions and time
step separations, to find out how much these simulation and other differences
affect observable properties in this model, and how well the different
variations match current observations at several redshifts. 

Bimodality in the star formation rate occurred for all
implementations, with the scatter in the star  
forming branch roughly due to changes in accretion history, and the
center and scatter for the quiescent branch put in by hand (which
would otherwise be at zero star formation rate).
The division between quiescent and star forming galaxies as a function
of stellar mass and star formation rate differs from that found in 
\citet{Mou13}, but some difference may be due to differences stellar mass definitions.     
Another difference between these observations and the models is that 
the split between the model star forming and quiescent galaxies does not evolve as
strongly as redshift increases (to higher star formation rates).  
This is in part because
the center of quiescent branch, put in by hand, is currently fixed at all
redshifts, while observationally the quiescent branch increases in star
formation rate with redshift.  (The model star
forming
branch does rise, although not fast enough for its peak to
remain
within the star forming branch of \citet{Mou13} at high redshift.)

For stellar mass functions, no variant matched observations 
perfectly for all redshifts,
although the best fit to stellar mass functions 
at low redshift was based on a simulation with cosmology and
mass definition similar to that used to construct 
the model's star formation efficiencies.  Using a simulation based
upon the current best fit
cosmology
gives stellar mass functions which are too high at low redshift, consistent with the
halo mass function being larger in the current best fit cosmology.  
At low redshift, using the instantaneous mass as dark matter halo mass
rather than the smoothed and constrained $SAM\_M_{\rm vir}$ 
produced a slight increase in the stellar mass functions at the lower
stellar masses, and using both instantaneous mass and 
only subtracting the mass of the most massive progenitor
(rather than a weighted sum of all progenitor masses) gives a further
increase.  At high redshift, the stellar
mass functions seem less sensitive to
choices of halo mass and methods of inheriting stellar mass from
progenitors.  Some dependence upon cosmology remains.  All tend to be
high relative to the central value for the observed
stellar mass functions, especially near the bend in the stellar mass
function, although the models based upon the cosmology used to
calculate the star formation efficiencies might be argued to
be closer.

Dividing the quiescent and star forming galaxies according to
the bimodality seen in star formation rate, the quiescent galaxy
stellar mass function tended to be too high for all models at
 low stellar mass.
For the tree model, where satellite subhalos are assigned their infall 
mass aside from mergers,
satellites
are almost all quiescent (about 1 percent have enough random star formation
to be classified as star forming using the 'by eye' separation from the
$M^*-SFR$ diagrams).  The excess of faint quiescent galaxies
also appears in the other models, however, which can have 1/3 or more 
satellites which are forming stars.

The models had two variations for inheriting stellar mass, from
either 1 or 2 progenitors.  These two cases tend to bracket the
observed stellar mass function at low redshift and high stellar mass.
At higher redshift, they get closer to each other but also increase
relative
to the observed stellar mass function, with both eventually lying mostly above
at $z\sim 2$. 
The models all seem to be above the central stellar mass to halo mass
relations calculated from observations, although close enough to be
within errors for most cases.   Aside from the tree model,
the stellar mass to halo mass relation changed
significantly between the 1 and 2 progenitor variants at high stellar
mass, with the latter rising far above observations.

These results suggest some ways to improve the model.
\begin{itemize}
\item
Presumably the star formation efficiencies would work better for the
current cosmology simulations if the efficiencies 
were calculated assuming
the best current fit cosmology.  This should help with the overshoot
in
estimated stellar mass for many redshifts, perhaps including the
tendency to fall high at high redshift for all the models.
In addition, more data is now available on stellar mass functions and
star formation rates at these redshifts.

\item Tying the star formation rate of satellites to subhalo mass gain
  for
satellites is difficult in some halo finders, for instance those which
fix the satellite subhalo
mass to infall mass aside from mergers.  
Perhaps some other way of having the satellite
star formation rate evolve would also work (for instance the decaying
star formation rate in \citet{Lu14}, or the delayed quenching of
\citet{ajcf13}\footnote{I thank P. Behroozi for mentioning this latter
  variant is being explored by other groups.}).  

\item The quiescent branch, put in by hand, likely should
  increase in star formation rate as the redshift increases.

\item It seems a combination
of inheriting from 1 and 2 progenitors (or maybe more)
might succeed.  This could perhaps depend upon the stellar mass and/or mass ratios
of progenitors\footnote{I thank M. van Daalen for these suggestions.}.
For instance,
\citet{MosNaaWhi13} note that for higher halo mass, more stellar mass
comes in from mergers, so perhaps some sort of final halo mass
dependence would also be
appropriate.  The stellar mass to halo mass relation often shows a strong
sensitivity to 1 or 2 progenitors contributing stellar mass.

\item The high values of $M^*(M_h)$ relative to observations, if not
  improved by changing the cosmology to the current best fit 
(the current best fit cosmology has more high mass halos than that
used in the star formation efficiencies \citep{BolPla16}), may be
improved by making more stellar mass go into the ICL, even from a
single progenitor.

\end{itemize}

It would be also interesting if certain properties (e.g. the excess of
faint
quiescent galaxies at high redshift) could not be improved by only changing
the small number of physical assumptions currently in the model.
The underlying assumption is that galaxies self-regulate (e.g. as in
\citet{OWLS,HopQuaMur11}) their growth, so that the influx of halo mass
(and thus presumably baryons) combined with the mass dependence of star
formation is enough to capture many of the properties of evolution.

Looking at the ensemble of galaxy histories, with fixed final stellar
mass,
all 6 \citet{Bec15} based models considered had average histories differing from the semi-analytical
models.
For low final stellar mass galaxies, all 6 had very similar average
histories,
as final stellar mass increase, these became distinct.
For the highest final stellar mass studied, $M^* = 10^{11} M_\odot$,
the average histories of the 2
progenitor bol and bolp models, where the inherited stellar mass of a galaxy is the
sum of that of its two largest progenitors, follows the average
history of galaxies for which $M^*=M^*(M_h)$ at every time step.
Compared to the two semi-analytic models,
there is often more scatter around the average histories for the models
constructed here (except for the highest final stellar mass), 
and they require the same or more basis fluctuations to capture
at least 90\% of the variance.  It would be interesting to see how
other similar simple models, e.g., \citet{Wan07,Bou10,Cat11,MutCroPoo13,Lil13,TacTreCar13,Bir14,Lu14,Lu15}, compare in these ensemble properties as well.

Given the simplicity of this model, it seems interesting to pursue it
further, as it appears that many of its predictions work reasonably well at
redshifts above zero.  It would be interesting to look in further
detail at the simulated galaxy populations, for instance to see if
starbursting galaxies appear automatically within the population and
if so, if their number agrees with observations.
The dark matter simulations also include galaxy positions and
velocities, and the full cosmological cosmic web.
When more high redshift observations with large volume 
become available, it would be interesting to compare the model with additional
observations
 such as clustering
 a function of stellar mass, star formation rate (once that
is better understood in the model) and environmental properties.  For instance, if conformity of central
galaxies
in halos is indeed tied to halo growth \citet{HeaBehvdB16}, see also \citet{HahPorDek09}, 
this effect is built into this model, as are other properties tying 
stellar mass and star formation rate to halo growth.   It would also
be interesting to push the model to higher redshifts.

\section*{Acknowledgements}
I thank M. White for numerous discussions and use of his simulated
TreePM data set, and Phillip Harris for early calculations of galaxy
history PCA's.
I also thank M. Becker, P. Behroozi, L. Guzzo, J. Moustakas, A. Munoz, N. Padilla,
M. van Daalen, and F. van den Bosch,
for discussions,
A. Gonzalez and C. Mancone for help with and explanations of EZGAL, 
and J. Moustakas for sharing the PRIMUS results using different
stellar mass assumptions, which was very helpful in order to explore
their effects, and M. Becker, M. van Daalen and M. White for extremely helpful
suggestions on an earlier draft, and the referee for helpful
suggestions as well.
This work was performed in part at the Aspen Center for Physics which
is supported by National Science Foundation grant PHY-1066293.
I am also grateful to the ROE and the IAP 
for hospitality during this work, and to the ROE for the opportunity
to present this work in a seminar and get very helpful feedback as a result.
The TreePM simulation by M. White was run at NERSC, and many of the
models presented here were run there as well.
The Millennium Simulation databases used in this paper and the web application providing online access to them were constructed as part of the activities of the German Astrophysical Virtual Observatory (GAVO).
The Bolshoi-Planck simulation was performed by Anatoly Klypin within the Bolshoi project of the University of California High-Performance AstroComputing Center (UC-HiPACC) and was run at the NASA Ames Research Center.
JDC was supported in part by DOE.
\appendix
\section{How to make validation plots with your simulated data set}
\label{sec:howto}
The tests shown above are available (for testing one mock catalogue) at
https://www.github.com/jdcphysics/validation/ (the
code and data are in the code subdirectory vsuite).\footnote{To test the stellar mass function of more than one mock, modify the number
  of calls to the routine {\bf getsimstellar} in {\bf plot4tog} and
  {\bf plot4sep} .}  
They consist of a python
file, valid\_suite.py, and then several data files, many of which were
kindly made available by the people who made and analyzed the observations.
Observational data sets are described above in \S\ref{sec:obssmfs} and
Table \ref{tab:smfobs}.

To run the python codes, you also need to have matplotlib and numpy,
which will be imported when you run the program.
To produce the plots above in \S \ref{sec:simmeas} for one simulated model
the command is:

{\bf runsuite}(zcen, fname, hval, omm, slopeval,shiftval, \\ boxside, delz, ramin,  ramax,
decmin, decmax)

\medskip

\noindent Here are two examples:
\begin{itemize}
\item fixed time periodic box of side 256 $h^{-1}$ Mpc, redshift 0.45, data file
  ``inputfile.dat'', hubble constant 0.67, $\Omega_m = 0.31$, default
  split of \citet{Mou13}, i.e. Eq.~\ref{eq:primuscut}, between
  star forming and quiescent galaxies.

{\bf runsuite}(0.45,''inputfile.dat'',0.67,0.31,0,0,256,''perbox'')
\item light cone, where one wants to look at a slice $0.43 \leq z \leq
  0.47$, ra and dec both between $\pm 2$, otherwise same as above:

{\bf runsuite}(0.45, "inputfile.dat",0.67,0.31,0,0,-1,''lc'',0.02,-2,2,-2,2)
\end{itemize}
It can be useful to start with {\bf slopeval, shiftval} =0, and then
vary these by looking at the stellar mass-star formation rate diagram,
until the solid line falls between the two regions of star formation.

\noindent Parameters in detail:
\begin{itemize}
\item {\bf zcen:} central redshift
\item {\bf fname:} input file name, in quotation marks, i.e. ``inputfile.dat''
\item {\bf hval:} hubble constant $h$
\item {\bf omm:} $\Omega_m$
\item {\bf slopeval:} Adjustments to quiescent vs. star forming galaxy
  split 
of PRIMUS \citep{Mou13}, i.e. in Eq.~\ref{eq:primuscut}
\begin{equation}
\begin{array}{lll}
\log (SFR_{\rm min})&=& - 0.49+(0.65 + {\bf slopeval}) (\log M^*
-10)\\ &&
+1.07 (z - 0.1)+{\bf shiftval}
\end{array}
\label{eq:aprimuscut}
\end{equation}
divides quiescent and star forming galaxies.
\item {\bf shiftval:} see above ({\bf slopeval})
\item {\bf boxside:} box side in $h^{-1} Mpc$ for fixed time periodic
  box, any negative number for light cone
\item {\bf runname:} a string, for name of run, e.g. ``bolshoi''
\item {\bf delz:} $\delta z$, i.e. for light cone, keep galaxies in
  range $z-\delta z$ to $z+\delta z$, ignored for fixed time box.
\item {\bf ramin:} RA minimum value for light cone, ignored for fixed
  time box.
\item {\bf ramax:} RA maximum value for light cone, ignored for fixed
  time box.
\item {\bf decmin:} DEC minimum value for light cone, ignored for fixed
  time box.
\item {\bf decmax:} DEC maximum value for light cone, ignored for fixed
  time box. 
\end{itemize}

The input file is a list of galaxies in the simulation, in ASCII
format.  Each galaxy is a different row, and the order and units of
the entries are:

$\log_{10} M^* [M_\odot]$, SFR ($M^*[M_\odot]$/yr), RA, DEC, zred, ifsat, $\log_{10}M_h [M_\odot]$

\begin{itemize}
\item RA and DEC are ignored for a fixed time box, zred is used to
  calculate $M^*(M_h)$.
\item ifsat = 0 if a central, 1 if a sat
\item $M_h$ ideally is $M_{vir}$, units are $M_\odot$ (no $h$) 
\end{itemize}

Here is an example of part of an input file for a periodic box (ra and dec are both
set to 1 since they are  not used, and the redshift is set to that of
the box):
\begin{equation}
\begin{array}{ccccccc}
9.428e+00 &  7.236e-01 &1. & 1. & 0.00 &0 & 11.3704 \\ 
1.024e+01  & 1.913e-01 &1. & 1. & 0.00 &1 & 11.8632 \\
9.501e+00 &  7.360e-02 &1. & 1. & 0.00 &0 & 11.3944 \\
1.069e+01  & 1.459e-01 &1. & 1. & 0.00 &1 & 12.2967 \\
9.400e+00  & 1.365e-01 &1. & 1. & 0.00 &0 & 11.3559 \\
9.514e+00  & 7.470e-02 &1. & 1. & 0.00 &0 & 11.3944 \\
1.053e+01  & 1.760e-01 &1. & 1. & 0.00 &0 & 12.1996 \\
9.620e+00  & 1.136e-01 &1. & 1. & 0.00 &1 & 11.5112 \\
\end{array}
\end{equation}
The observational data used for the stellar mass functions depends
upon the redshift, for the stellar mass to halo mass relation the
curves are from \citep{MosNaaWhi13,BWCz8}, and the stellar
mass-star formation rate plots can be compared to Fig.~1 of
\citet{Mou13}, for example.

\section{$M^*(M_h)$ fits to Observations}
\label{sec:appmsmh}

 The observational relations for $M^*(M_{h})$ found by
\citet{MosNaaWhi13} are tuned to the observations of
\citet{PerGon08}
(664  arcmin${}^2$) and \citet{San12} (33 arcmin${}^2$).  Those by
\citet{BWCz8} have the observations of \citet{PerGon08} (664 arcmin${}^2$)
and
\citet{Mou13} (5.5 $deg^2$) in the same region.  Both sets of observational 
stellar masses are found using
the \citet{BC03} stellar population synthesis models with a Chabrier
IMF.
The
relation from \citet{MosNaaWhi13} is
\begin{equation}
\frac{M^*}{M_h} = 2 N /[(\frac{M_h}{M_1})^{-\beta}
+(\frac{M_h}{M_1})^{\gamma} ]
\label{eq:mosmsmh}
\end{equation}
(with their equations 11,12,13,14)
\begin{equation}
\begin{array}{ll}
\log M_1(z) &= 11.590  + 1.195 (1-a) \\
N(z) &= 0.0351 -0.0247 (1-a)  \\
\beta(z) &= 1.376-0.826(1-a) \\
\gamma(z) &= 0.608 +0.329(1-a) \\
\end{array}
\end{equation}
the best fit parameters are in their Table 1.
The 1-$\sigma$ errors on coefficients
for
$\log M_1,N,\beta,\gamma$ are (in order appearing above) are
( 0.236, 0.353, 0.0058, 0.0069, 0.153, 0.225, 0.059, 0.173).
Here $M$ is halo mass, $M_{200c}$, and $M^*$ is stellar mass.

The \citet{BWCz8} fit is of the form (their equation 3)
\begin{equation}
\begin{array}{lll}
\log_{10}M^*(M_h) & =& \log_{10} (\epsilon M_1) + f(\log_{10}(M_h/M_1)) -
  f(0) \\
& & \\
f(x) &=& - \log_{10}(10^{\alpha x} +1) \\
& & + \delta \frac{
         (\log_{10}(1+e^x))^\gamma}{1+exp(10^{-x})} \\
\label{eq:behmsmh}
\end{array}
\end{equation}
with parameters (section 5 of their paper)
\begin{equation}
\begin{array}{lll}
    \nu &= &e^{-4a^2} \\
    \alpha     & = &-1.412 + 0.731(a-1)\nu \\
    \delta      &= & 3.508 + (2.608(a-1) -0.043 \, z)\nu \\
   \gamma    &  = & 0.316 + (1.319(a-1) +0.279\, z)\nu \\
     \log_{10} M_1   &   = &11.514 + (-1.793(a-1)  -0.251\, z)\nu \\
     \log_{10} \epsilon & = &-1.777 + (-0.006(a-1)  -0.000 \, z)\nu \\
                              & & -0.119(a-1) \; .
\end{array}
\end{equation}
Both stellar mass to halo mass relations are plotted for each stellar
mass bin in Fig.~\ref{fig:msmh}, but errors are not shown.

\end{document}